\documentclass[letterpaper]{article} %
\usepackage{aaai2026}  %
\usepackage{times}  %
\usepackage{helvet}  %
\usepackage{courier}  %
\usepackage[hyphens]{url}  %
\usepackage{graphicx} %
\usepackage{natbib}  %
\usepackage{caption} %
\usepackage[T1]{fontenc}
\usepackage{bbold}
\usepackage{dsfont}
\usepackage{amsmath}
\usepackage{xspace}
\usepackage{multirow,enumitem}
\usepackage{graphicx}
\usepackage{amssymb}
\usepackage{enumitem}
\usepackage{svg}
\usepackage[ruled,linesnumbered]{algorithm2e}
\usepackage[capitalize,noabbrev]{cleveref}
\usepackage{caption}
\usepackage{subcaption}
\usepackage{amsthm}
\usepackage{booktabs}
\usepackage{colortbl}
\usepackage{xcolor}

\newsavebox{\largestimage}

\newcommand{\our}{\textsc{Off-FSP}\xspace}
\newcommand{\osp}{\textsc{Off-SP}\xspace}

\definecolor{colorfirst}{rgb}{.866,.945, 0.831} %
\definecolor{colorred}{rgb}{.972, 0.8, 0.8} %
\newcommand{\cellred}{\cellcolor{colorred}}
\newcommand{\textred}{\colorbox{colorred}}

\theoremstyle{plain}
\newtheorem{theorem}{Theorem}[section]
\theoremstyle{definition}
\newtheorem{definition}[theorem]{Definition}

\usepackage[textsize=tiny]{todonotes}

\nocopyright
\title{Offline Fictitious Self-Play for Competitive Games}

\author{
    Jingxiao Chen\textsuperscript{\rm 1},
    Weiji Xie\textsuperscript{\rm 1},
    Weinan Zhang\textsuperscript{\rm 1}\thanks{Correspondence to: Weinan Zhang and Ying Wen.},
    Yong Yu\textsuperscript{\rm 1},
    Ying wen\textsuperscript{\rm 1}\footnotemark[1]
}
\affiliations{
    \textsuperscript{\rm 1}Shanghai Jiao Tong University \\
    {\tt\small \{timemachine, wnzhang, ying.wen\}@sjtu.edu.cn}
}

\usepackage{bibentry}

\begin{document}

\maketitle

\begin{abstract}
Offline Reinforcement Learning (RL) enables policy improvement from fixed datasets without online interactions, making it highly suitable for real-world applications lacking efficient simulators.
Despite its success in the single-agent setting, offline multi-agent RL remains a challenge, especially in competitive games.
Firstly, unaware of the game structure, it is impossible to interact with the opponents and conduct a major learning paradigm, self-play, for competitive games. Secondly, real-world datasets cannot cover all the state and action space in the game, resulting in barriers to identifying Nash equilibrium (NE).
To address these issues, this paper introduces \our, the first practical model-free offline RL algorithm for competitive games.
We start by simulating interactions with various opponents by adjusting the weights of the fixed dataset with importance sampling.
This technique allows us to learn the best responses to different opponents and employ the Offline Self-Play learning framework. 
To overcome the challenge of partial coverage, we combine the single-agent offline RL method with Fictitious Self-Play (FSP) to approximate NE by constraining the approximate best responses away from out-of-distribution actions.
Experiments on matrix games, extensive-form poker, and board games demonstrate that \our achieves significantly lower exploitability than state-of-the-art baselines. Finally, we validate \our on a real-world human-robot competitive task, demonstrating its potential for solving complex, hard-to-simulate real-world problems.

\end{abstract}

\section{Introduction}

Multi-agent reinforcement learning (MARL) provides a powerful learning framework to tackle the problems in multi-agent systems and has been applied to a wide range of domains, such as Go~\cite{silver2017mastering}, strategy games~\cite{vinyals2019grandmaster}, robotics~\cite{yu2023asynchronous}, unmanned aerial vehicle~\cite{yun2022cooperative}, and network routings~\cite{ye2015multi}.
MARL typically relies on extensive interaction and exploration with accurate and efficient environments, hindering its application in the real world.
In many real-world multi-agent problems, 
such as football~\cite{kurach2020google}, negotiation~\cite{yang2020improving},  and crowdsourcing~\cite{gerstgrasser2021crowdplay}, the absence of reliable simulators makes it expensive and inefficient to interact with the environment and collect new data for training MARL agents.
Moreover, in special problems, such as wildlife protection~\cite{fang2016deploying}, learning MARL agents online by trial and error is unsafe, which could lead to danger for patrols or wildlife.
In these cases, learning from existing data can be extremely valuable as it does not require additional sampling. 
Offline MARL offers an excellent alternative to solve these issues by improving policies from previously collected datasets without further interactions~\cite{tseng2022offline,yang2021believe}.

\begin{figure}[t]
    \centering
    \includegraphics[width=0.48\textwidth]{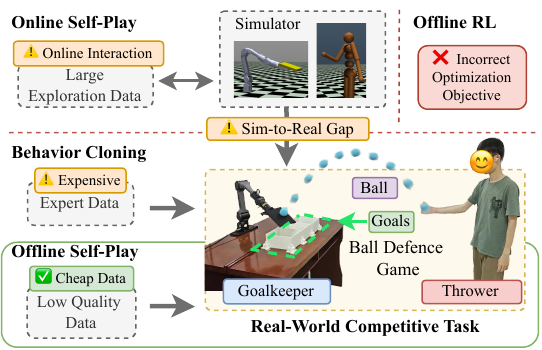}
    \caption{\textbf{Comparison of offline self-play with other learning paradigms for real-world competitive tasks.} Online self-play requires a simulator and suffers from the sim-to-real gap. Behavior cloning needs costly expert data. Naive offline RL optimizes with incorrect objectives.
    In contrast, \osp learns from inexpensive, low-quality data. See \Cref{sec:real_world} for experiments on the illustrated real-world game.}

    \label{fig:overview}
\end{figure}

In competitive multi-agent games, often framed as zero-sum games, the objective is to find a Nash equilibrium (NE)~\cite{Kreps1989} or maximizing the ELO ratio~\cite{silver2017mastering,ye2020towards}, which is divergent from the goal of maximizing cumulative rewards in single-agent situations.
Exemplified by studies of Texas hold'em poker~\cite{brown2018superhuman,brown2019superhuman}, the pursuit of a Nash equilibrium enables agents to learn robust policies against various opponents, enhancing their performance in competitive environments.
To achieve the goals, online approaches predominantly rely on the self-play paradigm~\cite{zhang2024survey}, wherein policies are continuously refined to maximize returns against evolving adversary policies.
However, in the offline scenario, the challenge arises from the absence of online interactions with evolving opponents, complicating the development of self-play paradigms.
Recent benchmarks of offline competitive games, such as AlphaStar Unplugged~\cite{mathieu2023alphastar} and Hokoff~\cite{qu2023hokoff,wei2022honor}, have directly applied single-agent offline RL algorithms with no self-play.
However, maximizing the cumulative rewards against static opponents results in an overfitting policy and vulnerable exploitation by more dynamic opponents. Moreover, when datasets are collected by poor policies, such as random policies, it is pointless to learn to defeat these weak opponents, because real-life opponents typically exhibit rational and high-performance behaviors.

The reliance on high-quality or high-coverage datasets, which are commonly expensive and suboptimal in the real world, is another challenge for offline learning.
The existing data-driven method, supervised learning, also called behavior cloning, ignores the objective of online learning and only imitates the sampling policy of datasets.  Consequently, this method performs poorly when dealing with non-expert datasets. Many recent works are learning to improve policies toward the NE, but they demand high state-action coverage within the dataset.
\citet{li2022offline} proposes a model-based offline paradigm for equilibrium finding, but it requires a strong assumption that datasets fully cover the state-action space of the original games.
\citet{cui2022offline, zhong2022pessimistic,cui2022provably} contribute to theoretical insights for a weaker assumption on datasets but only propose theoretically feasible algorithms. 
Existing work lacks a practical method applicable to non-expert and partially covered real-world datasets. 
\Cref{fig:overview} compares our paradigm with existing works, highlighting the advantage of \osp on solving real-world competitive problems.
Detailed comparisons with related works are provided in the \Cref{sec:related_works}.

In this paper, we propose an offline learning framework, called \textbf{Offline Self-Play (\osp)}, and an offline learning algorithm, \textbf{Offline Fictitious Self-Play (\our)}, for equilibrium finding in zero-sum extensive-form games, bridging the gap between single-agent offline RL and competitive games. To our knowledge, \our is the first model-free offline algorithm for practical zero-sum games
that offers the flexibility to combine with various Offline RL agents and improve policies on non-expert real-world datasets.
We first propose a technique to approximate interaction with different opponents by re-weighting the datasets with importance sampling. This allows us to learn the approximate best responses against arbitrary opponents with offline RL and derive a self-play paradigm under the offline setting.
Also, for partially covered and non-expert datasets, we use a surrogate loss, NashConv, to measure the distance to NE, rather than finding the exact NE.
\our combines single-agent offline reinforcement learning methods with fictitious self-play to learn approximate best responses iteratively and minimize the NashConv.
We validate our approach across matrix-form and extensive-form zero-sum games, demonstrating consistently low exploitability and superior performance over existing offline RL baselines under partially covered datasets. Notably, we further apply our method to a real-world human-robot competitive task, showcasing its potential in addressing hard-to-simulate decision-making problems.

\section{Preliminaries}

\subsection{Extensive-form Game}

\textbf{Extensive-form games} are a model of sequential interaction involving $n$ agents. Each player's goal is to maximize his payoff in the game. At each step $t$ of extensive-form game, only one player observes his respective \textbf{information states} $s^i_t \in \mathcal{S}^i$ and suggests his \textbf{action} $a^i_t \in \mathcal{A}^i(s^i_t)$.
$\mathcal{S}^i$ is the set of information states of player $i$, and $\mathcal{A}^i(s)$ is the set of available action at state $s$.
The \textbf{player function} $\mathcal{P}: \mathcal{S}\rightarrow\mathcal{N}$, with $\mathcal{N}=\{1,\dots, n\}$ denotes the set of players, determines the player to act.

In extensive-form games, each player plays following a \textbf{policy} $\pi^i: \mathcal{S}^i \rightarrow \Delta(\mathcal{A}^i)$ that maps information states to distributions of actions.
The \textbf{realization-plan} \cite{von1996efficient}, $x(s_t^i) = \Pi_{j=1}^{t-1} \pi^i(a_j^i | s_j^i)$, describes the probability of reaching the information state $s_t^i$ following player $i$'s policy, $\pi^i$.
The \textbf{strategy profile} $\pi = (\pi^1, \dots, \pi^n)$, is a joint of all player's policy. $\pi^{-i}$ denotes the strategy profile of all players excepts $i$. 
The \textbf{payoff} of player $i$ is denoted by $R^i(\pi^i, \pi^{-i}) \in \mathbb{R}$, and $\sum_i R^i =0$ for zero-sum games.
Given a fixed $\pi^{-i}$, \textbf{best response} $\textsc{BR}(\pi^{-i})$ is the policy with the highest payoff.
A \textbf{Nash equilibrium} (NE) is a strategy profile that any policy $\pi^i$ in this profile is a BR to the opponent's profile $\pi^{-i}$.
The \textbf{$\epsilon$-best response} ($\epsilon$-BR) and \textbf{$\epsilon$-Nash equlibirium} ($\epsilon$-NE) are approximations to the above definition.
$\epsilon$-BR is suboptimal by no more than $\epsilon$ compared with BR.
Similarly, $\epsilon$-NE is a profile of $\epsilon$-BR. \textbf{\textsc{NashConv}}~\citep{timbers2022approximate}, also called exploitability in two-player zero-sum games, evaluates the distance from $\pi$ to an NE, defined as $\sum_i R^i(\textsc{BR}(\pi^{-i}), \pi^{-i}) - R^i(\pi^i, \pi^{-i})$.

\subsection{Fictitious Self-Play}

\label{sec:fsp}
\textbf{Fictitious Self-Play (FSP)}\citep{heinrich2015fictitious} is a game-theoretic model that iteratively computes the best responses to opponents' average policy and updates their set of policies. 
We briefly describe FSP at ~\Cref{app:algo}.
In extensive-form games, the average policies $\pi^i_{k}$ are updated by the realization-equivalence theorem. 
At the $k$-th iteration of FSP, the average policy $\pi^i_{k}$ of player $i$ is
\begin{align}
\label{equ:avg_strategy}
\begin{split}
    \forall s, a: ~&\pi^i_{k}(a|s) = (1-\lambda)\pi^i_{k-1}(a|s) + \lambda \beta^i_{k}(a|s),\\
    \lambda &= \frac{\alpha_k x_{\beta^i_{k}}(s)}{(1-\alpha_k)x_{\pi^i_{k-1}}(s) + \alpha_k x_{\beta^i_{k}}(s)}, 
\end{split}
\end{align}
where $\beta^i_{k} \in \epsilon_k\textsc{-BR}(\pi_{k-1}^{-i})$ is a best-response to opponent $\pi_{k-1}^{-i}$ and $\alpha_k$ is the mixing parameter.
Policy $\pi^i_{k}$ is equivalent to choosing either policy $\pi^i_{k-1}$ or $\beta^i_{k}$ before the beginning of each game, with probabilities $1-\alpha_k$ and $\alpha_k$ respectively. 
A standard choice is $\alpha_k = \frac{1}{k}$.

\begin{figure*}[thbp]
    \centering
    \begin{minipage}{0.6\linewidth}
        \centering
        \includegraphics[width=\textwidth]{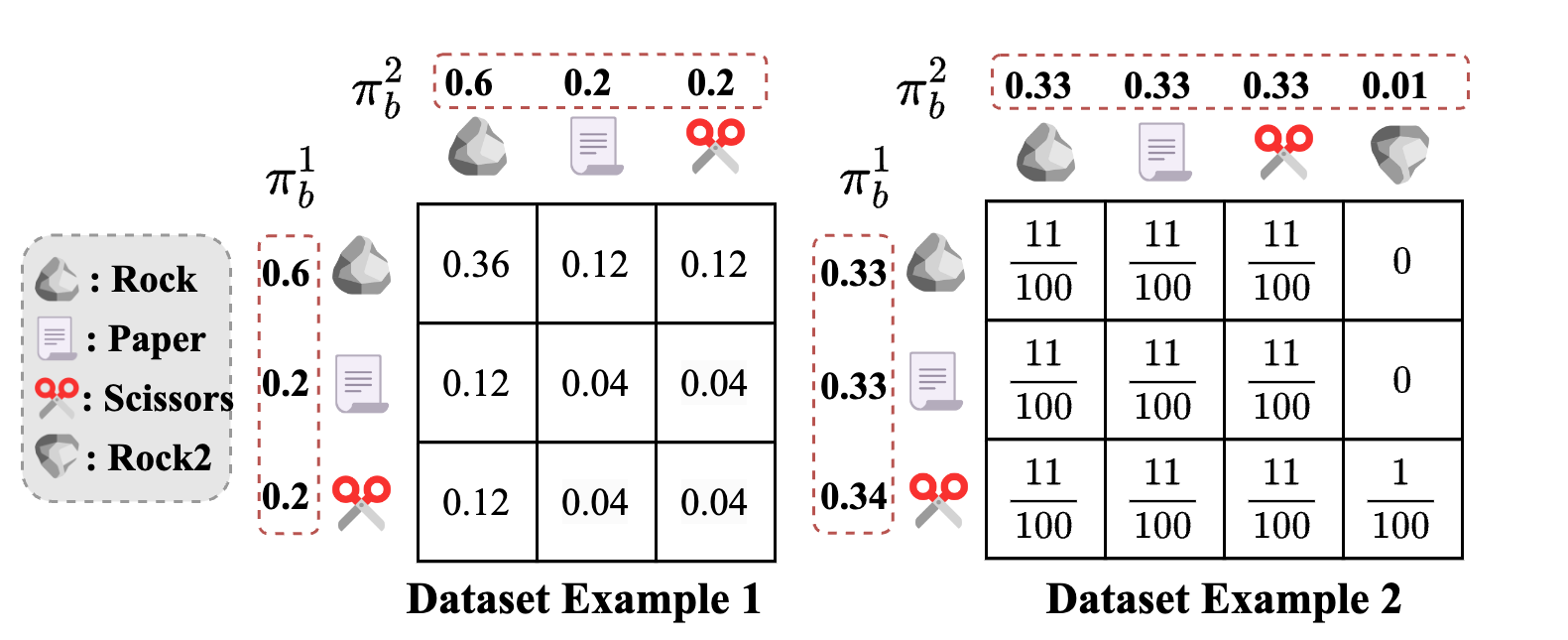} 
    \caption{\textbf{Example Datasets of RPS.} Numbers in the grids show the probability density of different samples. The red dashed boxes indicate the probability of different actions for corresponding behavioural policies.}
      \label{fig:rps_example}
    \end{minipage}
    ~~~
    \begin{minipage}{0.3\linewidth}
        \centering
            \includegraphics[width=\linewidth]{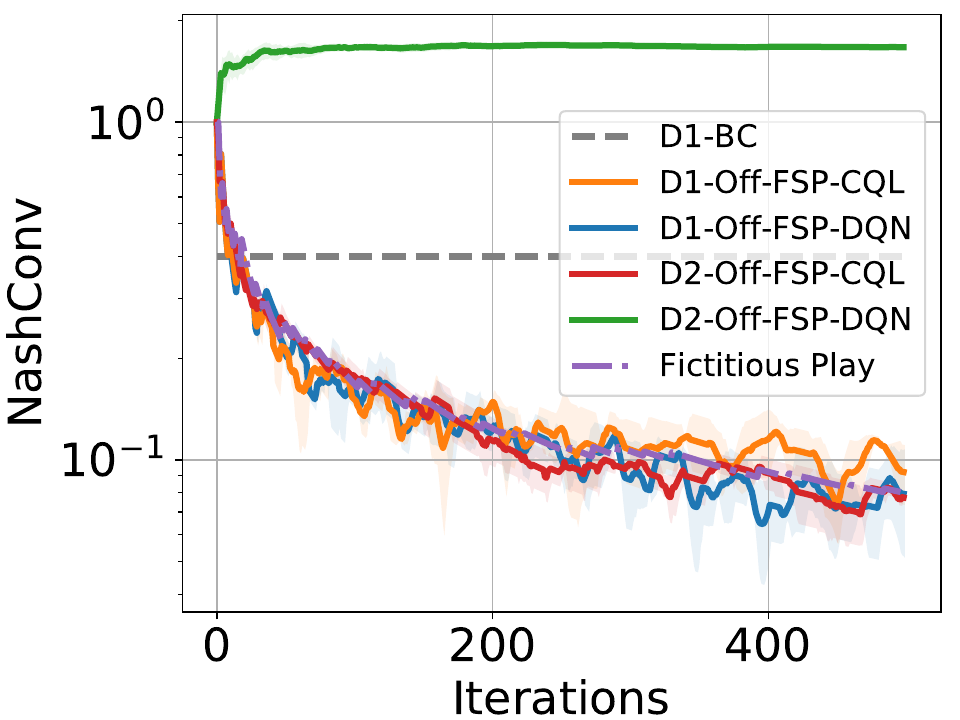}
        \caption{\textbf{Results of RPS.} The prefix of D1- and D2- are refering to restuls on the first and second datasets respectively.}
        \label{fig:rps_result}
    \end{minipage}
   \vspace{-1em}
\end{figure*}

\subsection{Offline Reinforcement Learning}

\textbf{Reinforcement learning} (RL) agents aim to maximize the expected cumulative discounted reward in a Markov decision process (MDP). MDP is denoted as a tuple $\mathcal{M} = (\mathcal{S}, \mathcal{A}, \mathcal{T}, \rho_0, r, \gamma, T)$. $\mathcal{S}, \mathcal{A}$ represent the state and action spaces. $\mathcal{T}(s_{t+1}|s_t, a_t), r(s_t, a_t)$ represent the dynamics and reward function. $\rho_0$ is the distribution of initial state $s_1 \sim \rho_0$. $\gamma\in [0, 1]$ is the discount factor.
The expected cumulative discounted reward following policy $\pi$ can be formalized as the action-value function $Q(s_t,a_t) = \mathbb{E}_{\pi} [\sum^{\infty}_{i=t} \gamma^{i-t} r(s_i,a_i)]$. %
Given the fixed opponent $\pi_{-i}$ and an extensive form game, the game of player $i$ can be defined as a MDP $\mathcal{M}(\pi_{-i})$~\cite{silver2010monte, greenwald2013solving}.
An $\epsilon$-optimal policy of the MDP, $\mathcal{M}(\pi^{-i})$, is also the $\epsilon$-BR to the policy $\pi^{-i}$.

\textbf{Offline RL} algorithm breaks the assumption that the agent can interact with the environment. The offline RL algorithm learns to maximize the expected cumulative reward based on a fixed dataset. 
The dataset
is sampled following a behavior policy $\pi_b(a|s)$ in MDP $\mathcal{M}$.
$\hat \pi_b(a|s) := \frac{\sum_{s,a\in\mathcal{D}^i} \mathds{1}[s=s,a=a]}{\sum_{s\in\mathcal{D}^i} \mathds{1}[s=s]}$ denote the empirical behavior policy, at all state $s\in\mathcal{D}^i$.
The Q-function may be erroneously overestimated at out-of-distribution (O. O. D.) actions, which is called extrapolation error~\cite{fujimoto2019off}.
Recent offline RL methods encourage the policy to learn on the support of training data or employ weighted behavior cloning to mitigate this error.

\section{The Motivating Example: Rock-Paper-Scissors}
\label{sec:rps_motivation}
We start by analysing the well-known game, Rock-Paper-Scissors (RPS), to illustrate the challenges of learning in offline datasets of competitive games. In this game, two players can choose one of three actions: \underline{R}ock, \underline{P}aper, or \underline{S}cissors. The payoffs are defined as follows: Rock beats Scissors, Scissors beats Paper, and Paper beats Rock. The game has a unique Nash equilibrium (NE) where both players play each action with equal probability of $\frac{1}{3}$. In \Cref{fig:rps_example}, we show two datasets of RPS. The first dataset is a fully covered dataset sampled from a non-expert policy, $P(\text{R,P,S}) = (0.6, 0.2, 0.2)$. The second dataset is a partially covered dataset sampled from a modified asymmetric RPS game, where the second player has a new action, Rock2, with the same payoff as Rock. 

In the first non-expert dataset, neither Behavioral Cloning (BC) nor naive offline RL can find the NE, which highlights the importance of the self-play paradigm.
 BC mimics the distribution of the dataset by supervised learning. Its policy, $P_{\text{BC}}(\text{R,P,S}) = (0.6, 0.2, 0.2)$, is suboptimal and beaten by the policy of playing Paper only. 
As the solution of ~\citet{mathieu2023alphastar,qu2023hokoff}, single-agent offline RL maximizes the payoff under the fixed opponent in the dataset, and its policy, $P_{RL}(\text{R,P,S}) = (0, 1, 0)$, is also easy to be exploited by another policies. Under the self-play paradigm, the policy learns to play against a dynamic opponent, which is more robust and can approximate the NE.
\Cref{fig:rps_result} compares the performance of our method, \our-CQL, with the baselines, BC, \our-DQN and online Fictitious Play (FP)~\cite{brown1951iterative}. 

Corresponding to the challenge of learning in partially covered datasets, the second dataset in \Cref{fig:rps_example} leaves the payoff of Rock2 not fully observed.
In such datasets, finding the accurate NE of the original game is impossible~\citet{}, so the goal of learning is to minimize the exploitability and NashConv, i.e., the distance to NE.
In the dataset, the only observed sample of Rock2 is $(\text{Scissors}, \text{Rock2})$. In player $2$'s perspective, the Rock2 action gets a payoff of $1$ all the time, so choosing it with a probability of $1$ is the best choice.
Nevertheless, the policy has the highest exploitability in the original game, as player $1$ possesses a best response by selecting scissors, resulting in a payoff of $1$ for player $1$.
Offline RL algorithms, such as Conservative Q-Learning(CQL)~\cite{kumar2020conservative}, mitigate this problem by punishing the agent for learning OOD actions, including unseen actions and undersampled actions.  
By combining self-play and offline RL, \our ignores the samples with large uncertainty and learns robust policies with low exploitability.
As shown in ~\Cref{fig:rps_result}, \our without offline RL (\our-DQN) converges to high exploitability, while \our with CQL can reduce the exploitability to a low level. 
We also show the learning curve of an online algorithm, FP, which shows the convergence speed of \our is comparable to online algorithms.
Further explanations of the results are given in the \Cref{sec:exp_rps}.

\begin{figure*} \centering
    \begin{subfigure}{0.24\textwidth}
        
        \centering
        \raisebox{2. em}{
            \includegraphics[width=\linewidth]{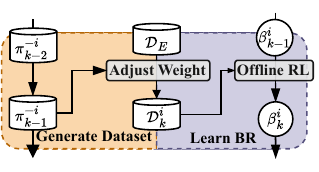}}
        \caption{Offline Self-Play framework.}
        \label{fig:off_sp}
    \end{subfigure}
    \begin{subfigure}[b]{0.35\textwidth}
        \includegraphics[width=\linewidth]{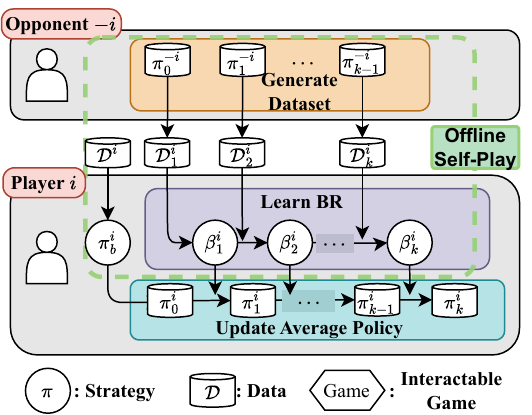}
    \caption{Pipeline of \our.}
        \label{fig:offline_fsp}
    \end{subfigure} %
    \begin{subfigure}[b]{0.39\textwidth} 
    
        \includegraphics[width=\linewidth]{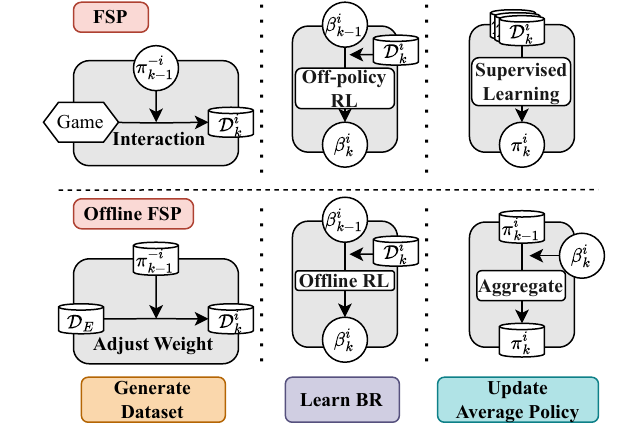}  
    \caption{Differences between FSP and \our. }
        \label{fig:fsp_diff}
    \end{subfigure} 
    \caption{Illustration of \osp, \our and three essential steps. The green box in (b) is \osp.} 
    \vspace{-1.5em}
        \label{fig:fsp_total}
\end{figure*}

\section{Offline Fictitious Self-Play}

In this section, we introduce the Offline Self-Play (\osp) learning framework and give an implementation as Offline Fictitious Self-Play (\our) algorithm to minimize NashConv, the distance to NE, with a fixed dataset.
\osp learns policies iteratively, maximizing cumulative rewards against changing opponents without interactable environments. 
\our adopts the fictitious self-play on \osp, where the policy plays against the average of past opponents.

Initially, we describe the offline datasets and make assumptions to simplify the problem.
Based on the original FSP, as described in \Cref{app:algo}, we derive the offline fictitious self-play with modifications on three essential functions, \textit{GenerateData}, \textit{LearnBestResponse}, and \textit{UpdateAveragePolicy}. In \osp, \textit{GenerateData} simulates play against different opponents with weighted datasets, and \textit{LearnBestResponse} optimizes the policy with an existing single-agent offline RL algorithm. 
Offline RL algorithm, such as CQL, ensures the performance of BR and reduction of NashConv even on partially covered datasets.
In \Cref{sec:calc_avg_strategy}, we introduce \textit{UpdateAveragePolicy} by computing the average policy on samples and derive \our. 

\subsection{Problem Formulation}
\label{sec:dataset}

    The trajectory of extensive-form games is a sequence of information states, actions, and rewards. 
    The trajectory is
    $\tau_E = \{s_1, a_1, r_1, \dots, s_T, a_T, r_T\}$.
    In player $i$'s perspective, the extensive-form game can be modeled as an MDP $\mathcal{M}$ given a fixed opponent $\pi^{-i}$~\cite{silver2010monte,greenwald2013solving}. In MDP $\mathcal{M}$, the dynamic function $\mathcal{T}(s^i_{t+1} \mid s^i_t, a^i_t)$ models the dynamics of opponents.
    
    The goal of \our is to iteratively learn the best-response policy $\pi^i$ against changing opponents and minimize NashConv.
    To learn best-response with single-agent RL, we project the trajectory $\tau_E$ into player $i$'s perspective with a projection function $\tau^i = \mathcal{F}^i(\tau_E)$. 
    The projection $\mathcal{F}^i$ filters out the states corresponding to player $i$ while relabeling the time indices.
    \begin{align*}
        \begin{split}
            \tau^i = \mathcal{F}^i(\tau_E) &= \{s_t, a_t, r_t \mid \forall s_t \in \tau_E, \mathcal{P}(s_t)= i\} \\
        & = \{s_1^i, a_1^i, r_1^i, s_2^i, \dots, s_{T'}^i, a_{T'}^i, r_{T'}^i\},
        \end{split} 
    \end{align*}
    where $T'$ is the length of the player $i$'s trajectory $\tau^i_{\mathcal{M}}$.
    
The subscripts of $s_t$ and $s_t^i$ are different and represent the time indices in the corresponding trajectories. 
    For state \( s^i_t \in \tau^i_{\mathcal{M}} \), we use function $I(s^i_t)$ to denote the subscript of state in $\tau_E$, and \( \tau_{<}^j(s_{t}^i) \) denotes opponent \( j \)'s latest state at \( s_t^i \).
    \begin{align*}
        \tau_{<}^j(s_{t}^i) = s_{k}, ~\text{where}~{k = \max\{ k' < I(s^i_t) \mid p_{k'} = j \}}.
    \end{align*}

\begin{figure}[ht]
  \begin{center}
    \centerline{\includegraphics[width=0.8\linewidth]{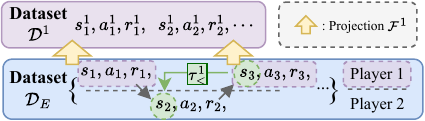}}
    \caption{An illustration example of trajectory $\tau^1$ and $\tau_E$. Purple parts represent Player 1. The yellow arrows imply the projection relationship under function $\mathcal{F}^1$. Green part indicates $\tau_{<}^1$.
    }
    \label{fig:dataset_example}
    \end{center}
    \vspace{-2em}
\end{figure}

In order to offer a clear explanation of the stated components, we present an example trajectory in Figure~\ref{fig:dataset_example}.
Based on the description of trajectory in both extensive-form games and single-agent perspective, the datasets consist of multiple corresponding trajectories. For extensive-form game, the dataset is $\mathcal{D}_E = \{\tau_E\}$. The dataset for player $i$ is $\mathcal{D}^i = \{\tau^i_{\mathcal{M}} = \mathcal{F}^i (\tau_E) \mid \tau_E \in \mathcal{D}_E\}$.
The $i$-th player's dataset, derived from dataset $\mathcal{D}_E$, is denoted as $\mathcal{D}^i=\mathcal{F}^i(\mathcal{D}_E)$.

To introduce the theoretical foundation of our method, we begin by defining an idealized dataset condition that facilitates analysis. These assumptions are not required for practical application and are used only to isolate and clarify the core mechanisms of our approach. In later sections, we show how the method generalizes to realistic datasets with partial coverage by integrating standard offline RL algorithms.

\begin{definition}[Fully Covered Dataset] \label{def:cover_d}$\mathcal{S} \text{~and~} \mathcal{A}$ represent the joint sets of information states and actions for all players $\mathcal{N}$.
    A dataset $\mathcal{D}_E$ in extensive-form games is a \textbf{fully covered dataset} if $\forall s \in \mathcal{S}, a \in \mathcal{A}(s) \Rightarrow (s,a) \in \mathcal{D}_E$.
\end{definition}

When a dataset is fully covered, the trained learning algorithm will not suffer from OOD problems. 
\begin{definition}[Real-Equivalence Dataset]
    For a dataset of extensive-form game $\mathcal{D}_E$ sampled following a policy $\pi_b$, dataset $\mathcal{D}_E$ is an \textbf{real-equivalence dataset} if 
        $Pr(\tau_E) = \frac{\sum_{\tau\in\mathcal{D}_E} \mathds{1}[\tau=\tau_E]}{|\mathcal{D}_E|}, \forall \tau_E \in \mathcal{D}_E,$
    where $Pr(\tau_E)$ is the probability of sampling trajectory $\tau_E$ with $\pi_b$.
\end{definition}
When a dataset is a real-equivalence dataset 
, sampling trajectories from the offline dataset $\mathcal{D}_E$ is equivalent to sampling from the extensive-form game with policy $\pi_b$ online and the right-hand side of the equation 
is the probability density of $\tau_E$ in $\mathcal{D}_E$, denoted by $\mathcal{D}_E(\tau_E)$.

\subsection{Offline Self-Play}
\label{sec:re-weight}

Self-play iteratively interacts with the game to generate data and utilizes the data to learn the best responses. In $k$-th iteration, the opponent for player $i$ is changed to $\pi^{-i}_{k-1}$, therefore, in order to learn the best response $\beta^i_k = \text{BR}(\pi^{-i}_{k-1})$, player $i$ must interact with $\pi^{-i}_{k-1}$ to generate new data $\mathcal{D}^i_k$.
The process of learning BR, from the perspective of player $i$, is equivalent to interacting with a new MDP $\mathcal{M}(\pi^{-i}_k)$ and maximizing returns with RL~\cite{greenwald2013solving}. 

In offline settings, we are limited to using a fixed dataset sampled by behavioural policy $\pi_b$, in which RL methods can only obtain the best response $\text{BR}(\pi_b^{-i})$ for player $i$. In order to execute the fictitious self-play, we generate player $i$'s datasets under different MDP $\mathcal{M}(\pi^{-i}_k)$.
With importance sampling~\cite{nachum2019dualdice}, we can emulate sampling from another dataset $\mathcal{D}^i_w$ with weighting $w(d_t) = \frac{\mathcal{D}^i_w(d_t)}{\mathcal{D}^i(d_t)}$~\cite{hong2023beyond} on the original dataset $\mathcal{D}^i$, where $\mathcal{D}^i_w(d_t)$ and $\mathcal{D}^i(d_t)$ denote the probability density of tuple $d_t=(s^i_t,a^i_t, s^i_{t+1})$. This formulation gives the following equivalence:
\begin{align}
\label{equ:importance_sampling}
    \mathbb{E}_{d_t\sim \mathcal{D}^i_w}[L(d_t;\theta)] \Leftrightarrow \mathbb{E}_{d_t\sim \mathcal{D}^i}[w(d_t)L(d_t;\theta)],
\end{align}
where $L(d_t; \theta)$ is the loss function of an off-policy RL method, and $\theta$ is the parameter of the RL policy to be optimized. 
With the assumption that dataset $\mathcal{D}_E$ is both a fully covered dataset and a real-equivalence dataset, sampling from it with importance sampling is equivalent to sampling from the online game with importance sampling. 
\begin{theorem}
\label{thm:reweight}
For player $i$, the weight of transferring the opponent from $\pi^{-i}_b$ to $\pi^{-i}$ is:
\begin{align}
\label{equ:weight}
    w(d_t) = \frac{x^{-i}_{\pi^{-i}}(s_j) {\pi^{-i}}(a_j| s_j)}{x^{-i}_{\pi_b^{-i}}(s_j) {\pi_b^{-i}}(a_j| s_j)}, ~
    s_j = \tau_{<}^{-i}(s_{t+1}^i).
\end{align}
\end{theorem}
The proof of Theorem~\ref{thm:reweight} is provided in Appendix~\ref{app:reweight_details}. Given an offline dataset $\mathcal{D}_E$, the policy $\pi_b$ can be approximated by an empirical behavioural policy $\tilde \pi_b$. 
We can estimate $\tilde \pi_b$ by counting or supervised learning policy.

In the learning process, the weight $w(d_t)$ assigned to a sample $d_t$ may shift to zero or a large value. 
As a result, the RL loss suffers from a large variance~\cite{munos2016safe}, leading to unstable training. To address this issue, we employ $\frac{w(d_t)}{\sum_{d\in\mathcal{D}}w(d)}$ as the sampling probability rather than multiplying the loss function by $w(d_t)$ as shown in Equation~\ref{equ:importance_sampling}. 
In this way, re-sampling from fully covered real-equivalence datasets is identical to sampling data in online games. With batches of data sampled from the dataset, we can apply an offline RL to learn BR.

\Cref{fig:fsp_diff,fig:off_sp}  illustrates these two functions and \osp. 
In the function \textit{GenerateData}, we first calculate $w(d)$ for all the samples $d_t$ and get a re-weighted dataset $\mathcal{D}_k^{i}$ for each player $i$.
In the function \textit{LearnBestResponse}, offline RL algorithms repeatedly sample a batch of data and optimize the learned policy for $M$ times. We use CQL~\cite{kumar2020conservative} as the default offline RL algorithm in our implementation, i.e., $\beta^i_k = \text{CQL}(\mathcal{D}_k^{i})$.
These two functions derive the \osp. At each step $t$, \osp first generates a dataset playing against the current opponents and updates policies towards the best response of the opponents.

\begin{figure*}[htbp]
    \centering
    \begin{minipage}[t]{0.66\linewidth}
        \centering
        \includegraphics[width=0.32\linewidth]{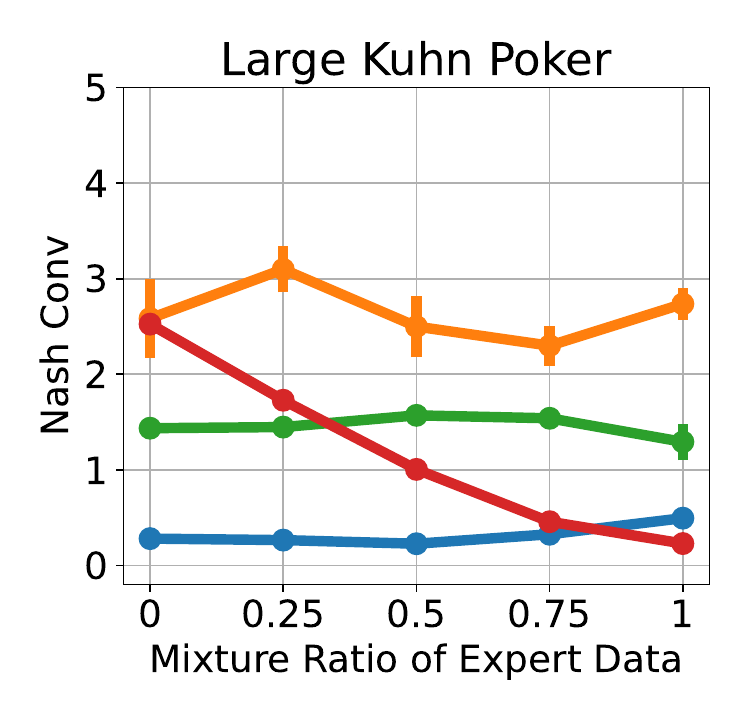}
        \includegraphics[width=0.32\linewidth]{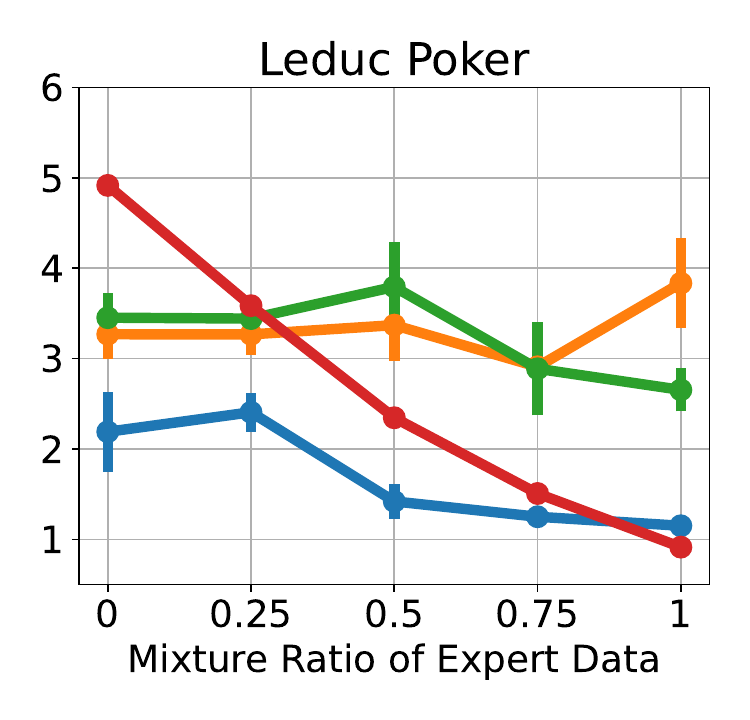}
        \includegraphics[width=0.32\linewidth]{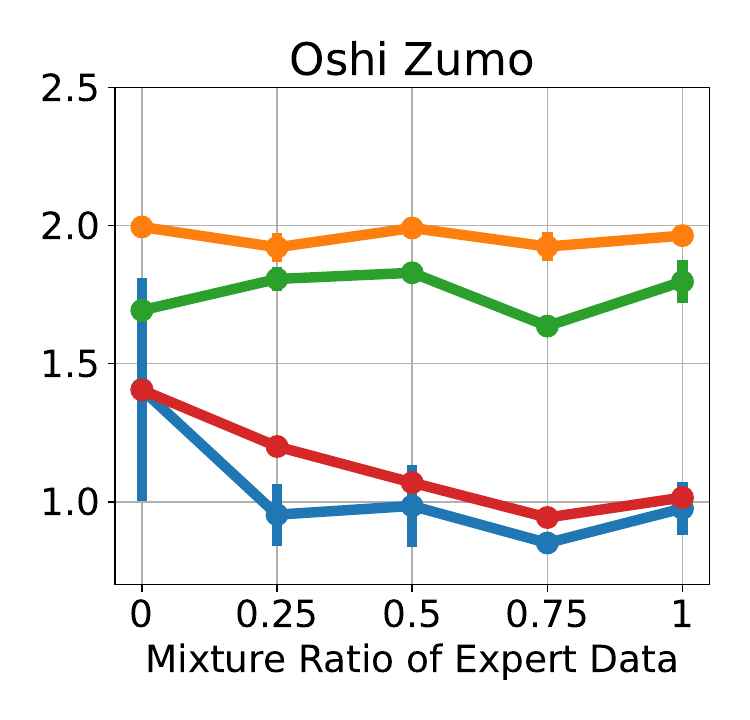}
        \\
        \includegraphics[width=0.32\linewidth]{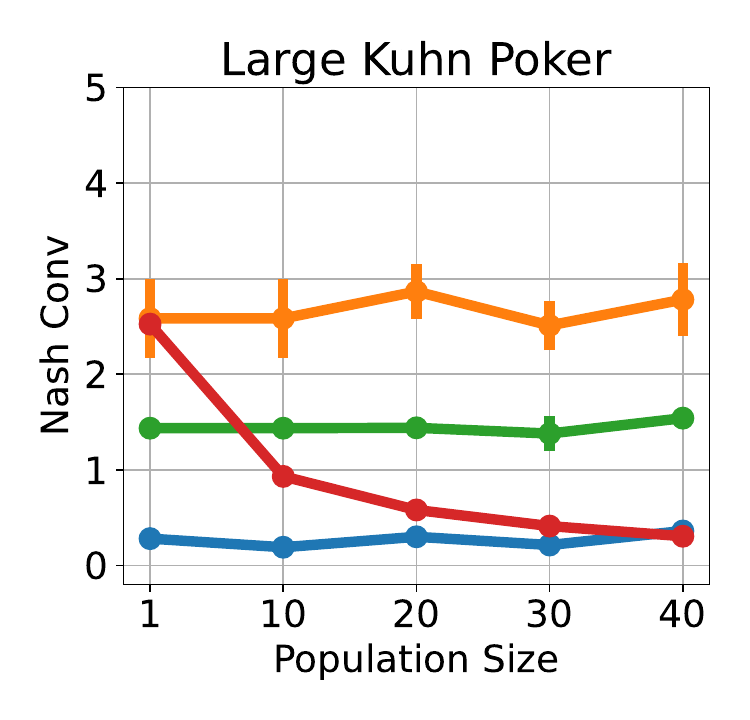}
        \includegraphics[width=0.32\linewidth]{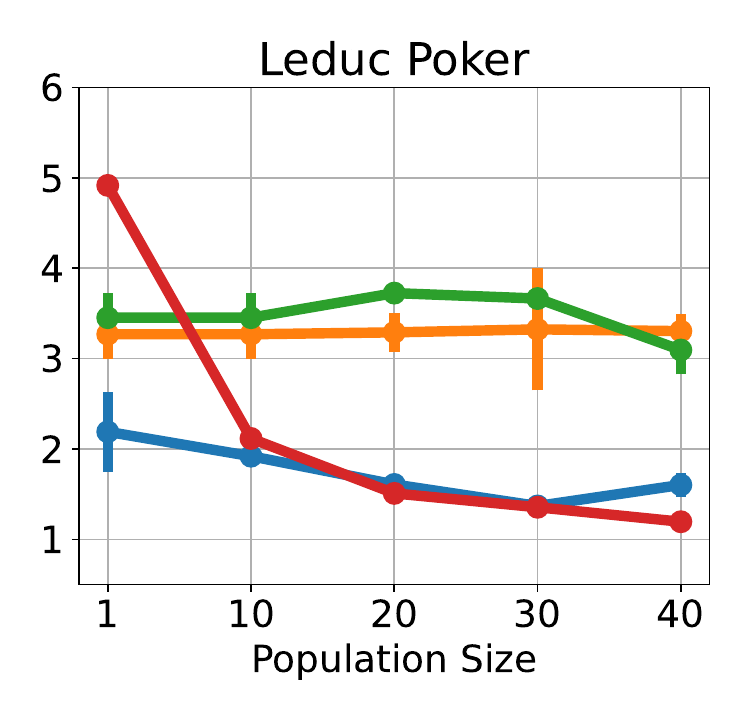}
        \includegraphics[width=0.32\linewidth]{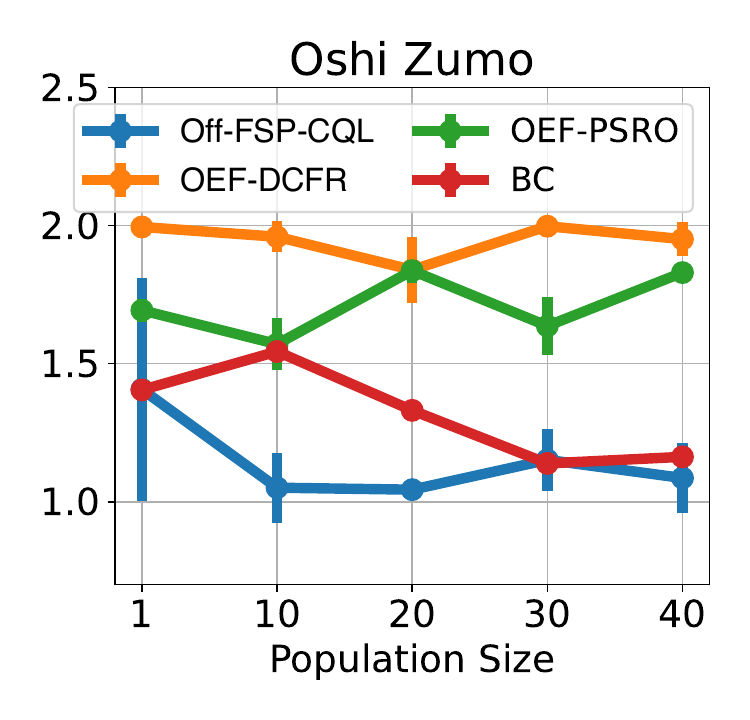}
        \caption{Results on Extensive-Form Games. \textbf{(Top)} NashConv on Mix Datasets; \textbf{(Bottom)} NashConv on Population Datasets.}
        \label{fig:extgame_nc}
    \end{minipage}
    ~~~
    \begin{minipage}{0.27\linewidth}
        \centering
        \vspace{15mm}
        \includegraphics[width=\linewidth]{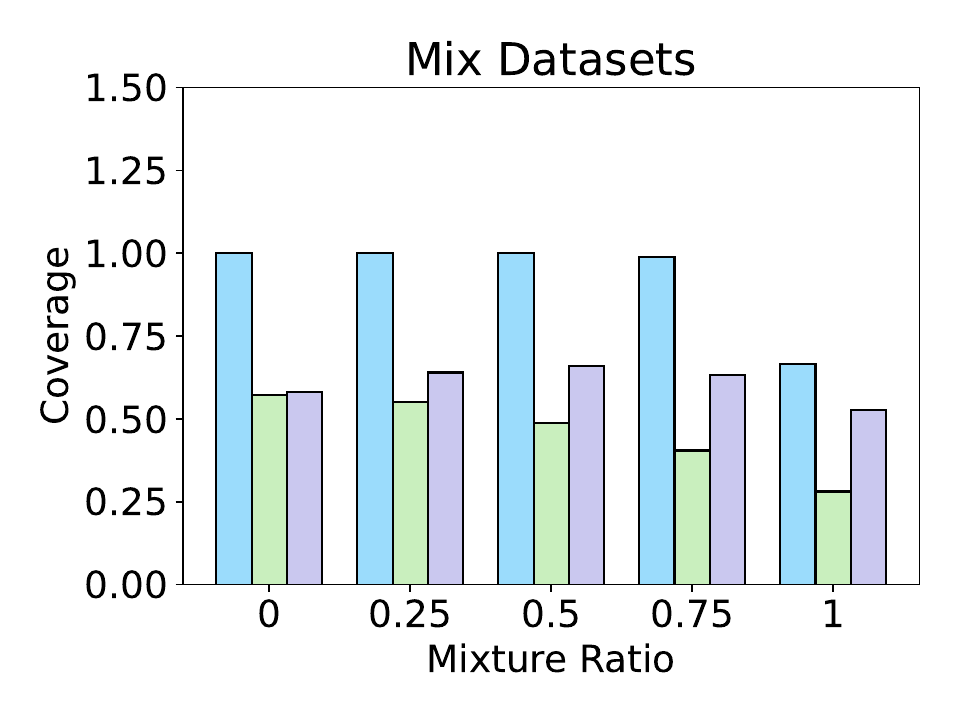}
        \\
        \includegraphics[width=\linewidth]{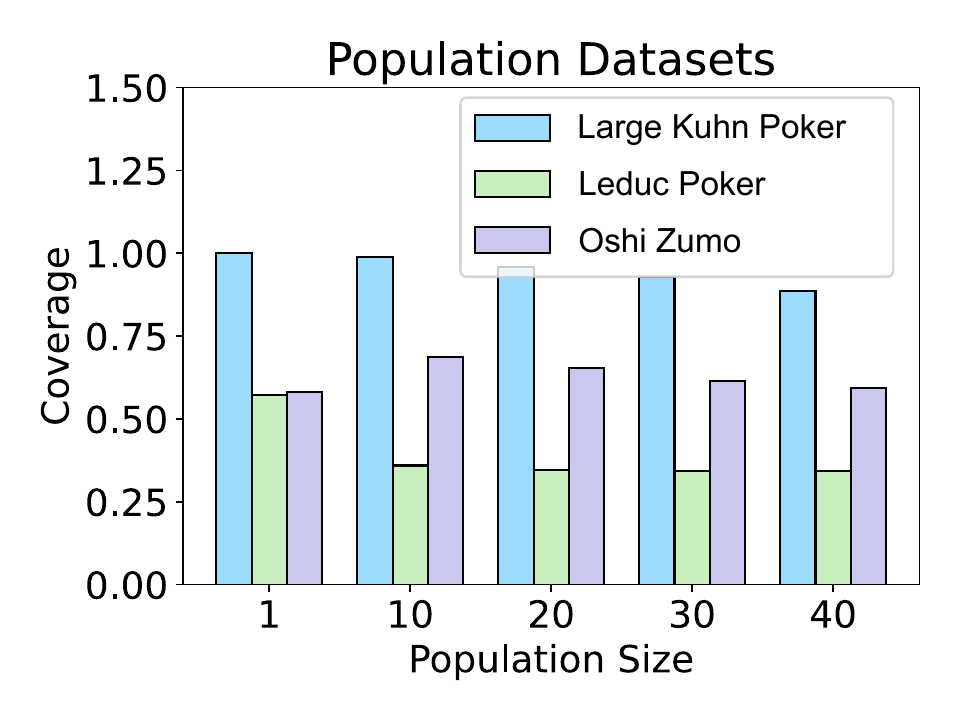}
        \vspace{-1em}
        \caption{Coverage ratio of datasets.}
        \label{fig:extgame_dataset}
    \end{minipage}
\end{figure*}

\subsection{Aggregate Average Policy in Samples}
\label{sec:calc_avg_strategy}

Similar to FSP, \our play against the average policy $\pi_{k+1}$ following ~\Cref{equ:avg_strategy}.
Although an interactable average policy $\pi_k$ is not necessary in the offline setting, we can just maintain the probability $\pi_k(s,a)$ in the datasets for the weighting technique. Traverse along the trajectory $\tau_i$, \our computes the probability $\pi^i_{k}(s,a)$ for player $i$ following~\Cref{equ:avg_strategy}. \Cref{fig:fsp_diff} shows one step of the function \textit{UpdateAveragePolicy}. To facilitate the calculation, we save the probability into the current dataset $\mathcal{D}^i_k$, changing the sample into $d'=(s, a, \pi^i_{k}(s,a))$. In short, we have:
$$
 \pi^i_{k}(s,a) \leftarrow\frac{k-1}{k} x^{i}_{\pi^{i}_{k-1}}(s)\pi^i_{k-1}(s,a) + \frac{1}{k}  x^{i}_{\beta^{i}_{k}}(s)\beta^i_{k}(s,a).
$$

In the evaluation phase, we aggregate all policies within a collection $\Pi$. Before the beginning of each game, one of the policies in $\Pi$ is chosen with a specific probability, and the payoff of the aggregation is the expected payoff over all possible policies.
To keep the storage efficient, we only keep policies at fixed interval steps for the real-world application in \Cref{sec:real_world}.
Algorithm~\ref{alg:off-fsp} in \Cref{app:algo} and \Cref{fig:offline_fsp} presents the pipeline of Offline Fictitious Self-play. 

\noindent\textbf{Memory and Computational Complexity.} 
Compared with offline RL, \our only assigns additional weights for samples in the dataset, so the memory complexity is still $O(|\mathcal{D}_E|)$ in training. 
The training time of \our can be noted as $T = T_{reweight} + T_{learn}$, where $T_{reweight}$ is the time of re-weighting and $T_{learn}$ is the time of offline RL learning. In empirical experiments of \Cref{sec:exp}, the $T_{reweight} : T_{learn} \approx 1 : 1$.

\section{Experiments} \label{sec:exp}
In this section, we design experiments to show the performance of \our in offline competitive games. 
Three benchmark extensive-form games are selected for analysis: Leduc Poker, Large Kuhn Poker, and Oshi Zumo, which include two stochastic poker games and one deterministic board game. These environments allow efficient sampling and easy NashConv computation, facilitating a comprehensive assessment of algorithm performance.
We compare \our with state-of-the-art baselines, including OEF~\cite{li2022offline} and Behavior Cloning (BC).
We further conduct ablations on offline RL components to demonstrate the flexibility of \our.
Finally, the practical applicability of \our is highlighted through deployment in a real-world human-robot competitive task, showcasing its potential for solving complex, hard-to-simulate problems. Further experimental details are provided in the appendix.

\subsection{Extensive-form Games}
\label{sec:exp_leduc}
To further evaluate \our on complicated extensive-form games, we collected datasets of different quality on Leduc Poker, Large Kuhn Poker, and Oshi Zumo. 
Following the previous work~\cite{li2022offline}, the first type of datasets is called mix datasets, which are sampled from a mixture of expert and random policies. Evaluation on random, mixed, and expert datasets is also a common practice in single-agent offline RL~\cite{ kumar2020conservative}. We sample 5 different mix datasets for each game, the ratio of sampling from the expert are $0, 0.25, 0.5, 0.75, 1$. The expert policy is learned by online PSRO~\cite{lanctot2017unified} with 40 iterations. The second type of dataset is called the population dataset, which is sampled from the population of online PSRO. $5$ population datasets are uniformly sampled from all population policies of $1, 10, 20, 30, 40$ iterations of online PSRO. This type of dataset is designed to simulate the datasets sampled from a population of people with different levels of expertise and is similar to the setting in the real world.
The population datasets with $1$ iteration and the mix datasets with a ratio of $0$ are random datasets. 
Every dataset comprises 10\,000 trajectories, which is far less than the number in the online PSRO.
We visualize the coverage of terminal states, i.e., leaf nodes, in \Cref{fig:extgame_dataset}. Most of the datasets are partially covered.

\Cref{fig:extgame_nc} shows the NashConv of \our and baselines on three extensive-form games. 
The difficulty of learning from datasets is affected by two factors: the quality of sampling policies and the coverage of datasets. NashConvs of BC show the quality of sampling policies. In most cases, \our shows the best performance, and the performance of \our is also robust to both the quality of sampling policies and the coverage of datasets. 
Both OEF-DCFR and OEF-PSRO, two variants of OEF~\cite{li2022offline}, fail in most cases.
OEF tries to find an offline equilibrium with the model-based paradigm, but it is easy to be misled by O.O.D. actions. OEF mixed its policy with BC, evaluated it in online games multiple times, and used the minimum NashConv as a result. To make a fair comparison, we removed the mixing operation in OEF from our main experiment. The results with mixing are shown in ~\Cref{app:more_results}.
Compared with BC, \our outperforms in non-expert datasets and achieves comparable performance in expert datasets. 
In expert datasets, the diversity of samples is limited, and few samples can be exploited to improve the policy. \our aims to improve the policy with non-expert datasets, which is more practical and easier to scale up in real-world problems.

\begin{figure}[htbp]
    \vspace{-1em}

    \centering

        \includegraphics[width=0.23\textwidth]{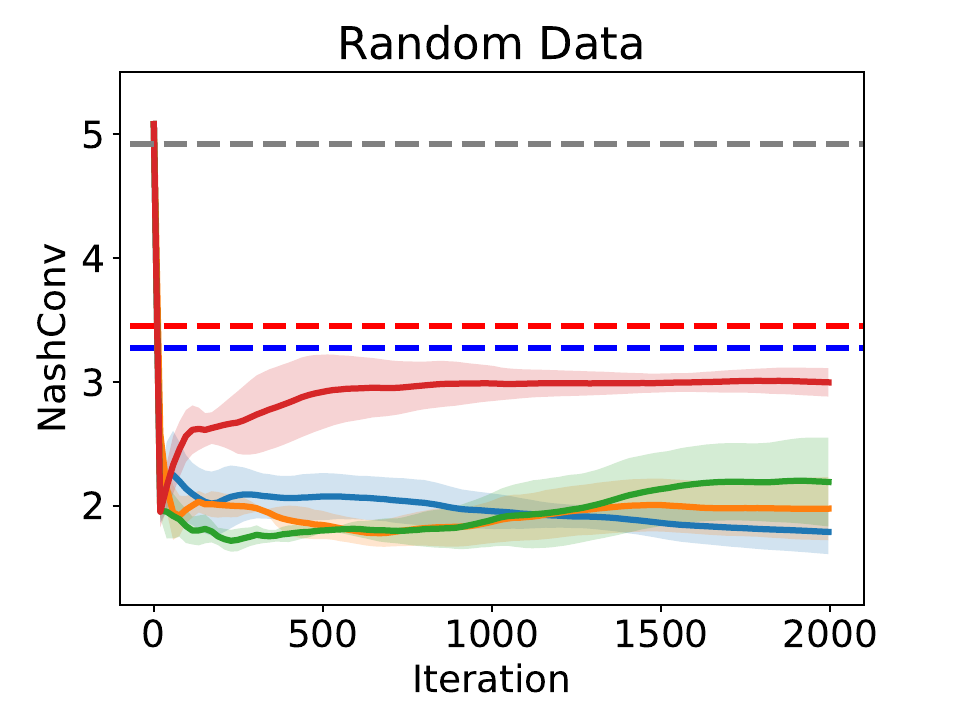}
        \includegraphics[width=0.23\textwidth]{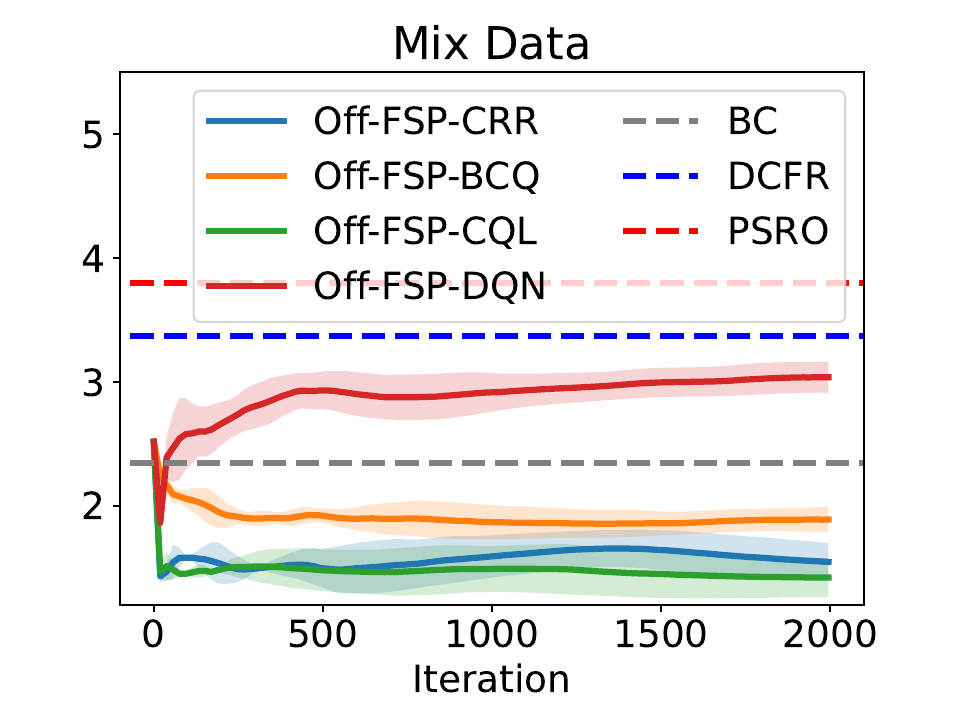}
        \caption{Results of ablation study on random and mix data of Leduc Poker.}\label{fig:ablation}
    \end{figure}

\noindent \textbf{Ablation Studies.}
In previous experiments, we showed the performance of \our with CQL as the offline RL algorithm to learn the best response.
However, the choice of an offline RL algorithm is also an important factor. 
\Cref{fig:ablation} shows learning curve of \our with different RL algorithms, including CQL, BCQ~\cite{fujimoto2019off}, CRR~\cite{wang2020critic}, and DQN on two datasets of Leduc Poker.
Without the constraint of offline RL to avoid O.O.D. actions, \our with DQN converges to a policy with high exploitability.
Corresponding to the design of \our, it has the potential to combine with any offline RL algorithms. \our with CQL, BCQ, and CRR show similar performances in these two datasets.

\subsection{Real-world Human-Robot Confrontation}
\label{sec:real_world}
\begin{figure}[t]
    \centering
    \includegraphics[width=0.48\textwidth]{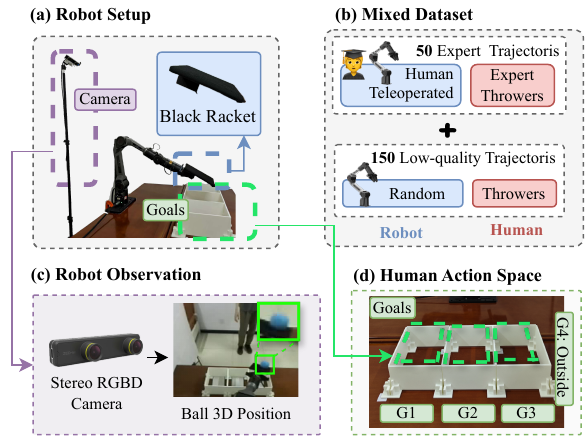}
    \caption{Setups of the human-robot competitive game. G1 to G4 in sub-figure (d) are four actions of the human player.}\label{fig:ball_setup}
\end{figure}

To further evaluate \our, we design a complex real-world zero-sum game—a human-robot ball defence task, which serves as a simplified version of football defense and attack. 
\Cref{fig:overview,fig:ball_setup} illustrate the task scenario.
This setup involves a human player, making it difficult to simulate and costly to collect data, thus highlighting the significance of \our.
The game consists of two players: a human thrower, who attempts to throw a ball into one of three baskets (goals), and a robotic arm acting as a goalkeeper that tries to block the ball. The human thrower receives a reward of $+1$ if the ball enters a goal and $-1$ otherwise. As a zero-sum game, the robot receives the negative of the human's reward and uses a black racket to defend the goals.
The robot is equipped with a stereo RGBD camera that estimates the ball’s position in real time (see \Cref{fig:ball_setup}c). Its observation space includes the ball's position and velocity, as well as the position of the robot's end effector. The robot policy outputs target end-effector positions at $60$~Hz. These are executed via inverse kinematics and a PD controller. 
For the human player, we simplify their action space to four discrete targets: the three baskets (G1 to G3) or outside the basket area (G4), as shown in \Cref{fig:ball_setup}d. The target is solely determined by the ball's flight trajectory after release and does not change whether the robot blocks it. Therefore, G4 corresponds to imprecise human throws and represents a poor strategy.
The setup of this task is shown in \Cref{fig:ball_setup}. 

We collected a mixed dataset comprising $50$ expert trajectories and $150$ low-quality trajectories. Expert data were collected with a human operator teleoperating the robot arm while another expert performed the throws. The average win rate of the teleoperated robot in these expert trajectories was $64\%$. 
Low-quality trajectories were collected by deploying the robot with a random policy, which moves to uniformly sampled positions within a constrained action space. The human thrower was not an expert and occasionally missed the targets (G4).
This protocol makes low-quality data cheaper to collect, requiring only a non-expert human thrower.

\begin{table}[htbp]
    \centering
    \begin{tabular}{l|c|c|c|c|c}
        \toprule
        \textbf{Methods} & \textbf{G1} & \textbf{G2} & \textbf{G3} & \textbf{Worst} & \textbf{Average} \\
        \midrule
        BC & 0.55 & \cellred 0.35 & 0.4 & 0.35 & 0.43 \\
        BC-Expert & \cellred 0.45 & 0.5 & 0.65 & 0.45 & 0.53 \\
        CQL & \cellred 0.4 & \textbf{0.85} & 0.8 & 0.4 & 0.68 \\
        \midrule
        \our-CQL & \textbf{0.8} & \cellred 0.75  & \textbf{0.9} & \textbf{0.75} & \textbf{0.82} \\
        \bottomrule
    \end{tabular}
    \caption{Winning rate of robot goalkeeper in ball defense game. The \textred{colored} numbers represent the worst case between goals.}
    \label{tab:ball_results}
\end{table}

Since one player is human, we evaluate only the robot's policy across different algorithms. As \textit{NashConv} is difficult to compute in real-world settings, we adopt the robot’s win rate as the evaluation metric. We compare \our-CQL with three baselines: BC, BC-Expert, and CQL. Here, BC-Expert denotes BC policy trained only on trajectories with a reward of $+1$. OEF failed in this task, so we removed it from the comparison.

For each goal, a total of 20 throws were performed collectively by 10 human throwers (2 throws per thrower on average). As shown in \Cref{tab:ball_results}, \our-CQL outperforms all baselines with consistently robust performance. 
The dataset shows an imbalanced distribution $P(\text{G1,G2,G3,G4}) = (0.29, 0.11, 0.32, 0.28)$, with relatively fewer throws to the middle goal (G2) and a high proportion of off-target throws (G4).
\our learns a more balanced human policy $(0.33, 0.22, 0.40, 0.04)$, leading to improved robustness against diverse human strategies.
By focusing only on high-reward data, BC-Expert and CQL overfit to average opponents in the dataset and fail to defend consistently across all goal targets (G1–G3).

\section{Conclusion}

In this paper, we study offline multi-agent reinforcement learning for competitive games and propose \osp and \our to enable single-agent offline RL algorithms to be applied in this scenario. We find that \our can approximate NE even with partially covered datasets. Extensive experiments show that all variants of \our significantly outperform state-of-the-art baselines in different datasets of multiple two-player zero-sum games.

  \bibliography{ref}

\newpage
\newpage
\appendix
\onecolumn
\setcounter{secnumdepth}{1}

\section{Algorithms} \label{app:algo}

\Cref{alg:fsp} describes Fictitious Self-Play (FSP) in the online setting. \Cref{alg:off-fsp} is the Offline Fictitious Self-Play.
Compared with FSP, the function \textit{GenerateData} and \textit{UpdateAveragePolicy} only require two additional traverses of the dataset and skip the time of interacting with the environment, so it does not introduce too much computational overhead.

\begin{algorithm}[H]
\caption{Fictitious Self-Play}
\label{alg:fsp}

\SetKwFunction{FSP}{FictitiousSelfPlay}
\SetKwFunction{GenerateData}{GenerateData}
\SetKwFunction{LearnBestResponse}{LearnBestResponse}
\SetKwFunction{UpdateAveragePolicy}{UpdateAveragePolicy}

\SetKwProg{Fn}{Function}{}{end}
\Fn{\FSP{}}{
    Initialize policy $\pi_0$.\;
    
    \For{$k=1$ \KwTo $K$}{
        
        $\mathcal{D}_k \leftarrow$ \text{\underline{GenerateData}}({$\pi_{k-1}$})
    
        \ForEach{player $i \in \mathcal{N}$}{
            $\beta^i_k \leftarrow$ \text{\underline{LearnBestResponse}}({$\mathcal{D}_k, \pi^{-i}_{k-1}$})
            
        }
        $\pi_k \leftarrow$ \text{\underline{UpdateAveragePolicy}}({$\pi^i_{k-1}, \beta^1_k, \dots, \beta^n_k$})
    }
    \KwRet{average policy $\pi_K$}\;
}
\end{algorithm}

\begin{algorithm}[H]
\caption{Offline Fictitious Self-Play }
\label{alg:off-fsp}

\SetAlgoLined
\SetKwFunction{FMain}{Off-FictitiousSelfPlay}
\SetKwFunction{FGenerateData}{GenerateData}
\SetKwFunction{FLearnBestResponse}{LearnBestResponse}
\SetKwFunction{FUpdateAveragePolicy}{UpdateAveragePolicy}
\SetKwProg{Fn}{Function}{:}{}
\SetKwInOut{Input}{input}
\SetKwInOut{Output}{output}

\Fn{\FMain{$\mathcal{D}_E$}}{
    Estimate the empirical behavioural policy $\tilde \pi_b$\;
    
    $\Pi_i \leftarrow \{\pi_b^i\}, \quad \pi_0 \leftarrow \tilde \pi_b$\;
    
    $\mathcal{D}^i_0 \leftarrow \{(s,a,\pi^i_0(s,a)) | (s,a) \in \mathcal{F}^i(\mathcal{D}_E)\}$\;
    
    \For{$k=1$ \KwTo $K$}{
        \For{each player $i\in \mathcal{N}$}{
            $\mathcal{D}^i_k \leftarrow$ \text{\underline{GenerateData}}({$\mathcal{D}_E, \mathcal{D}^{-i}_{k-1}$})
            
            $\beta^i_k \leftarrow$ \text{\underline{LearnBestResponse}}({$\mathcal{D}^i_k, \beta^i_{k-1}$})
            
            $\mathcal{D}^i_{k} \leftarrow$ \text{\underline{UpdateAveragePolicy}}({$\mathcal{D}^i_{k-1}, \beta_k^i$})
            
            Update collection of BRs $\Pi_i \leftarrow \Pi_i \cup \{\beta_k^i\}$\;
        }
    }
    \KwRet{ Aggregate($\Pi$)}
}
\BlankLine
\Fn{\FGenerateData{$\mathcal{D}_E, \mathcal{D}^{-i}_{k-1}$}}{
    Get probability of $\pi_{k-1}^{-i}$ from $\mathcal{D}^{-i}_{k-1}$\;
    
    Calculate $w_{\pi_{k-1}^{-i}}$ following Equation~\ref{equ:weight}\;
    
    $\mathcal{D}^i_{k} \leftarrow$ WeightData($\mathcal{D}_E, w_{ \pi_k^{-i}}$)\;
    
    \KwRet{$\mathcal{D}^i_{k}$}
}
\BlankLine
\Fn{\FLearnBestResponse{$\mathcal{D}^i_k, \beta^i_{k-1}$}}{
    Initialize policy parameters with $\beta^i_k \leftarrow \beta^i_{k-1}$\;
    
    Optimize $\beta^i_k$ with offline RL and data $\mathcal{D}^i_{k}$\;
    
    \KwRet{$\beta^i_k$}
}
\BlankLine
\Fn{\FUpdateAveragePolicy{$\mathcal{D}^i_{k-1}, \beta^i_{k}$}}{
    \For{$(s,a, \pi_{k-1}^i(s,a)) \in \mathcal{D}^i_{k-1}$}{
        Calculate $\pi_k^i(s,a)$ with $\beta^i_{k}, \pi^i_{k-1}$ following Equation~\ref{equ:avg_strategy}, and update dataset $\mathcal{D}^i_{k}$\;
        
    }
    \KwRet{$\mathcal{D}^i_{k}$}
}

\end{algorithm}

\section{Related Work}\label{sec:related_works}
In competitive games, also known as zero-sum games, people have developed a series of learning algorithms, where
finding Nash equilibrium with self-play is the major learning paradigm for this problem. 
Fictitious self-play (FSP)~\cite{heinrich2015fictitious} combines fictitious play (FP)~\cite{brown1951iterative} with self-play and provably converges to an NE in extensive-form games. 
Counterfactual regret minimization (CFR)~\cite{bowling2015heads} combines regret minimization with self-play, first solving an imperfect-information game of real-world scale.
DeepNash~\cite{perolat2022mastering} extends regularised Nash dynamics with RL and converges to an approximate NE in Stratego. 
Policy-Space Response Oracles (PSRO)~\cite{lanctot2017unified} learn to find NE by iteratively learning best responses to its policies' population, which can also be seen as a population-based self-play.
Some works~\cite{silver2017mastering,ye2020towards,yang2022perfectdou} of large-scale games aim for a higher winning percentage or returns and try to exploit opponents, not pursuing strict NE, still based on the self-play paradigm. 

In the offline learning paradigm, behavior cloning (BC) is the simplest form that directly imitates the sampling policy of datasets. The performance of BC is limited by the sampling policy in the dataset, while offline RL can surpass this limitation by maximizing returns.
Offline RL faces the challenge of the extrapolation error on out-of-distribution states and actions. Batch-constrained deep q-learning (BCQ)~\cite{fujimoto2019off} and conservative q-learning (CQL)~\cite{kumar2020conservative} mitigate this issue by constraining the gap between policy and data distribution. Critic regularized regression (CRR) \cite{wang2020critic} solve this problem by weighted behavior cloning. 

In offline competitive scenarios, BC is a viable algorithm, still limited by the quality of sampling policy in datasets.
\citet{mathieu2023alphastar,qu2023hokoff} directly learn policies by single-agent offline RL without self-play. This way also requires high-quality behavior policies in datasets and is easily overfitted to fixed opponents.
\citet{cui2022offline,zhong2022pessimistic,cui2022provably} propose theoretically feasible algorithms.
\citet{li2022offline} proposed a model-based learning framework, OEF, for extending online equilibrium-finding algorithms in offline scenarios. 
However, OEF only supports datasets with full coverage of state-action spaces, which is not realistic for real-world problems. Although previous theoretic methods allow for partially covered datasets, they cannot be practically applied to real-life problems. In this paper, \osp enables the self-play paradigm in offline MARL of competitive games. \our implement a practical algorithm to find NE and has the potential to learn on partially covered datasets by integrating any variants of single-agent offline RL algorithms.

\section{Derivation of Theorem}

\label{app:reweight_details}

\begin{proof}
    Now we derive the weight when changing the opponent to $\pi_{-i}$. Since we do not need to adjust player $i$’s policy, player $i$’s policy is still considered as $\tilde \pi_b^i$.
In player $i$'s perspective, the reach probability of a tuple $d_t=(s^i_t, a^i_t, s^i_{t+1})$ is 
\begin{align*}
    Pr(d_t | \pi^{-i}) = Pr(s^i_0) \pi^i(a^i_{t+1}|s^i_{t+1}) \cdot \Pi_{j=0}^{t} \pi^i (s^i_j,a^i_j)\mathcal{T}_{\pi^{-i}}(s^i_{j+1}| s^i_j,a^i_j),
\end{align*}
where $\mathcal{T}_{\pi^{-i}}$ is the transition function under MDP $\mathcal{M}(\pi^{-i})$.
So we have:
\begin{align*}
        w_{\pi ^{-i}}(d_t) &= \frac{Pr(d_t | \pi^{-i})}{Pr(d_t | \tilde\pi^{-i}_b)} \\
        &= \frac{Pr(s^i_0) \tilde \pi_b^i(a^i_{t+1}|s^i_{t+1})}{Pr(s^i_0) \tilde\pi_b^i(a^i_{t+1}|s^i_{t+1})} \cdot \frac{\Pi_{j=0}^{t} \tilde \pi_b^i (s^i_j,a^i_j)\mathcal{T}_{\pi^{-i}}(s^i_{j+1}| s^i_j,a^i_j) }{\Pi_{j=0}^{t} \tilde \pi_b^i (s^i_j,a^i_j)\mathcal{T}_{\tilde \pi_b^{-i}}(s^i_{j+1}| s^i_j,a^i_j)} \\
        &= \Pi_{j=0}^{t} \frac{\mathcal{T}_{\pi^{-i}}(s^i_{j+1}| s^i_j,a^i_j)}{\mathcal{T}_{\tilde \pi_b^{-i}}(s^i_{j+1}| s_j,a_j)}\\
\end{align*}

Considering $\mathcal{T}_{\pi^{-i}}(s^i_{j+1}| s^i_j,a^i_j)$ and $\mathcal{T}_{\tilde \pi_b^{-i}}(s^i_{j+1}| s_j,a_j)$, both of these two items are the multiplication of a series of transition probabilities, of which only the probability of opponents' policies $\pi^{-i}$ is different. So the above equation can be simplified to:

\begin{align*}
    w_{\pi ^{-i}}(d_t) &= \Pi_{j=0}^{t} \frac{{\pi^{-i}}(a_j| s_j)}{{\tilde \pi_b^{-i}}(a_j| s_j)} \\ 
    &= \frac{x_{\pi^{-i}}(s_j) {\pi^{-i}}(a_j| s_j)}{x_{\tilde \pi_b^{-i}}(s_j) {\tilde \pi_b^{-i}}(a_j| s_j)},
\end{align*}
where $s_j = \tau_{<}^{-i}(s_{t+1}^i)$.
\end{proof}

\section{Baselines}

\paragraph{Behaviour Cloning (BC).} Behaviour Cloning directly learns the probability distribution of actions conditioned on the observation from the dataset, hoping to recover the performance of the behavior policy used to generate the dataset.

\paragraph{Conservative Q-Learning (CQL)} Conservative Q-Learning \cite{kumar2020conservative} learns a conservative Q-function such that the expected value of a policy under this Q-function lower-bounds its actual value in order to alleviate overestimation of values induced by the distributional shift between the dataset and the learned policy. In practice, we solve such an optimization problem to update our Q-function, where the first term is the penalty for lower-bounding the Q-function, and the second term is the Bellman error. We use CQL based on the QRDQN.
\begin{align}
    \min _Q \alpha \mathbb{E}_{\mathbf{s} \sim \mathcal{D}}\left[\log \sum_{\mathbf{a}} \exp (Q(\mathbf{s}, \mathbf{a}))-\mathbb{E}_{\mathbf{a} \sim \hat{\pi}_\beta(\mathbf{a} \mid \mathbf{s})}[Q(\mathbf{s}, \mathbf{a})]\right]+\frac{1}{2} \mathbb{E}_{\mathbf{s}, \mathbf{a}, \mathbf{s}^{\prime} \sim \mathcal{D}}\left[\left(Q-\hat{\mathcal{B}}^{\pi_k} \hat{Q}^k\right)^2\right]
\end{align}

\paragraph{Critic Regularized Regression (CRR)} Critic Regularized Regression \cite{wang2020critic} handles the problem of offline policy optimization by value-filtered regression. It selectively imitates the dataset by choosing a function $f$ that is monotonically increasing in $Q_\theta$. 

\begin{align}
\underset{\pi}{\arg \max } 
\ \mathbb{E}_{(\mathbf{s}, \mathbf{a}) \sim \mathcal{D}}\left[f\left(Q_\theta, \pi, \mathbf{s}, \mathbf{a}\right) \log \pi(\mathbf{a} \mid \mathbf{s})\right]
\end{align}

\paragraph{Batch-Constrained deep Q-learning (BCQ)} Batch-Constrained deep Q-learning \cite{fujimoto2019off} uses a generative model to learn the distribution of transitions in the dataset, samples several actions from it in the current state, perturbates them, and chooses the one with the highest Q value. In this way, they tried to minimize the distance of selected actions to the dataset and lead to states similar to those in the dataset. In the experiments, we use a discrete BCQ variant introduced in \cite{fujimoto2019benchmarking}. 

\paragraph{Single-Agent Offline RL} In this set of baseline methods, each player independently learns a policy from the dataset using single-agent offline RL algorithms, including CQL, CRR, and BCQ. It's theoretically equivalent to figuring out the best response to the fixed opponent policy used to collect the dataset. This kind of method is also employed in \citet{qu2023hokoff} and \citet{mathieu2023alphastar}.

\paragraph{Offline Equilibrium Finding (OEF)} Offline Equilibrium Finding \cite{li2022offline} is a model-based framework to find the game equilibrium in the offline setting. They directly train a model to describe the game dynamics and apply online equilibrium-finding algorithms like PSRO and DCFR (a deep learning version of CFR) to compute equilibrium. They also combine the behavior cloning policy with the model-based policy for improvement, where they mix these two policies with different weights and evaluate each weight in the online game to find the best combination.

\section{Details of experiments}
\label{app:exp_details}
\subsection{Datasets}
\subsubsection{Datasets of Rock, Paper, Scissors}
D1 is randomly sampled from a policy, $\pi_b = (0.6, 0.2, 0.2)$. D2 is generated according to the exact proportion described in~\Cref{fig:rps_example}.
Every dataset is composed of 1000 trajectories.

\subsubsection{Datasets of Extensive-form Games}

The Nash Conv of expert policies are $0.86, 0.26, 1.01$ in Leduc Poker, Large Kuhn Poker, and Oshi Zumo, separately.
Every dataset comprises 10\,000 trajectories
 \textbf{The size of samples in datasets is far less than the number of samples required by online algorithms. }
Taking PSRO as an example, training BR requires about 10\,000 trajectories at each iteration, and additional samples are needed to evaluate the new BR.

\subsection{Rock, Paper, Scissors}
\label{sec:exp_rps}

As described in \Cref{fig:rps_example}, the first dataset, D1, is randomly sampled from a uniform policy, $\pi_b = (0.6, 0.2, 0.2)$, which is a fully covered dataset and an approximation of the ideal dataset. D2 are partially covered datasets, which are constructed following the probability. \Cref{fig:rps_result} shows the NashConv within $500$ iterations and the average policies of player $2$ after the end of training.

In D1 dataset, BC learns a suboptimal policy with an exploitability of $0.4$, while the exploitability of \our is less than $0.1$ after $500$ steps.
\our-DQN is the ablation variant of \our without the offline RL algorithm, to learn the best response. It
is easy to be misled by the OOD actions in D2.
\our-CQL is the default setting of \our, which successfully learns a robust policy with low exploitability in both D1 and D2 datasets. 

\subsection{Extensive-form Games}
\textbf{Leduc Poker} is a simplified version of two-player Texas holdem. Leduc Poker is a widely-used imperfect information normal-form game as a testbed for online algorithms.

\textbf{Large Kuhn Poker} is a variant of Kuhn Poker where an initial pot for each player is $5$. Each player has four actions: Fold, Check, Call, and Raise $1$ pot. Only in the first $8$ steps can players raise.

\textbf{Oshi-Zumo} is a deterministic board game in which two players repeatedly bid to push a token off the other side of the board~\citep{buro2004solving}.  The size of the board is $3$. The initial number of coins for each player is $4$.
The maximum game horizon is $6$.

\subsection{Ball Defence Game}
The robot arm in the ball defence game is Agilex PiPER, a 6-DOF arm. It is equipped with an RGBD camera, Zed mini, to perceive the position of the blue ball. We use color segmentation with 1344x376 resolution and 100 FPS to detect the ball. The size of each basket, i.e., the goals, is 20x17x10 cm, and the size of the black racket is 16x13 cm.

In data collection, we restricted the action space of the robotic arm's end effector to the area above the basket and discretized it into a 10-dimensional discrete space.  This increases the success rate of data collection and ensures that the robot's movements will not cause any damage to the experimental setup.

In the evaluation phase, we invited 10 human players to play the ball defence game and evaluate the performance of all methods. For fair comparison, each human player first has a 3-minute warmup to familiarize themselves with this game, and then plays with each method in a random order. For each method, we only retain the first two results that each person throws at each goal. The total number of evaluations for each goal and each method is 20.

\subsection{Variants of \our}
\textbf{\our-CQL, \our-CRR,} and \textbf{\our-BCQ} are different variants of our methods that integrate with different single-agent offline RL methods in the function of LearnBestResponse. We then introduce details about these Offline RL methods.

In CQL, we use a QRDQN as the base algorithm and a simple MLP as the Q network. Hence, the network's input is the observation vector, and the output has a dimension of the number of quantiles, multiplied by the number of actions. The hidden layer of MLP contains the same number of neurons. We use ReLU as the activation function between every linear layer.

In CRR, the actor and critic networks have the same structure, first a shared feature network, and then an MLP head. The shared feature network is an MLP with two layers, and the size of the second layer is the number of actions. All activation functions are ReLU. The CQL penalty term is also added to the CRR objective function for better performance.

In BCQ, the policy network and the imitation network have the same structure. They are also first a shared feature network and then an MLP head, which is exactly the same as CRR.

\begin{table}[htb]
    \centering
    \caption{Hyperparameters used in Leduc Poker}
    \label{tab:hyp_leduc}
    \begin{tabular}{l|l|l|l}
    \hline
        Hyperparameter & CQL & CRR & BCQ \\ \hline
        Hidden layer number & 4 & 5 & 6 \\ \hline
        Hidden layer size & 256 & 128 & 256 \\ \hline
        Learning rate & 0.0001 & 0.0001 & 0.0001 \\ \hline
        Adam $(\beta_1,\beta_2)$ & (0.9, 0.999) & (0.9, 0.999) & (0.9, 0.999) \\ \hline
        Training iterations  & 2\,000 & 2\,000 & 2\,000 \\ \hline
        Training steps per iteration  & 1\,000 & 1\,000 & 1\,000 \\ \hline
        Target Network update frequency & 100 & 100 & 100 \\ \hline
        Batch size & 1024 & 1024 & 1024 \\ \hline
        QRDQN number of quantiles & 100 & ~ & ~ \\ \hline
        CQL $\alpha$ (mix) & 2.0 & 0.1 & ~ \\ \hline
        CQL $\alpha$ (population)& 0.5 & 0.1 & ~ \\ \hline
        BCQ unlikely action threshold & ~ & ~ & 0.1 \\ \hline
        BCQ imitation logits penalty & ~ & ~ & 0.01 \\ \hline
        CRR improve mode & ~ & Exponential \\ \hline
        CRR $\beta$ & ~ & 1 & ~ \\ \hline
        CRR ratio bound & ~ & 20 & ~ \\ \hline
    \end{tabular}
\end{table}

\begin{table}[htb]
    \centering
    \caption{Hyperparameters used in Large Kuhn Poker, Oshi Zumo and ball Defence Game.}
    \label{tab:hyp_others}
    \begin{tabular}{l|l|l|l}
    \hline
        Hyperparameter & Large Kuhn Poker & Oshi Zumo & Ball Defence \\ \hline
        Offline RL algorithm & CQL & CQL & CQL \\ \hline
        Hidden layer number & 4 & 4 & 4 \\ \hline
        Hidden layer size & 256 & 256 & 256 \\ \hline
        Learning rate & 0.0001 & 0.0001 & 0.0001 \\ \hline
        Adam $(\beta_1,\beta_2)$ & (0.9, 0.999) & (0.9, 0.999) & (0.9, 0.999) \\ \hline
        Training iterations  & 1\,000 & 1\,000 & 200 \\ \hline
        Training steps per iteration  & 1\,000 & 1\,000 & 50 \\ \hline
        Target Network update frequency & 100 & 100 & 10 \\ \hline
        Batch size & 1024 & 1024 & 128 \\ \hline
        QRDQN number of quantiles & 100 & 100 & 100 \\ \hline
        CQL $\alpha$ & 0.1 & 0.01 & 0.1 \\ \hline
    \end{tabular}
\end{table}

\subsection{Hyperparameters}

For OEF baseline, all settings are the same as \citet{li2022offline}.

In RPS, we use Adam optimizer with a learning rate of 0.005 and betas (0.9, 0.999), and the network has only one layer, and the number of quantiles in QRDQN is 100. The target network is not used. In each iteration, the network is trained for five epochs and is updated 100 times each epoch with a batch size of 1000.

In extensive-form games, the hyperparameters are shown in \Cref{tab:hyp_leduc,tab:hyp_others}. All networks are optimized by Adam Optimizer \cite{kingma2017adam}. The number of hidden layers in CRR and BCQ describes the whole network, including the feature network and MLP head. The target network is used. Our experiments run on AMD EPYC 7302 CPU and 3080Ti GPUs.

In the ball defence game, the hyperparameters are shown in \Cref{tab:hyp_others}. To accelerate the training process, we use softmax with temperature 0.5 to calculate the probabilities of each action. In the final evaluation, we only keep a checkpoint of the robot policy at 100, 125, 150, 175, and 200 iterations and randomly pick one of them each time.

\section{Explanation on RPS experiments}
As described in \Cref{fig:rps_example}, the first dataset, D1, is randomly sampled from a uniform behavioural strategy, $\pi_b = (0.6, 0.2, 0.2)$, which is a fully covered dataset and an approximation of the ideal dataset. D2 are partially covered datasets, which are constructed following the probability. \Cref{fig:rps_result} shows the NashConv within $500$ iterations and the average strategies of player $2$ after the end of training.

In D1 dataset, BC learns a suboptimal strategy with an exploitability of $0.4$, while the exploitability of \our is less than $0.1$ after $500$ steps.
\our-DQN is the variant of \our by using DQN, an online RL algorithm, to learn the best response. The algorithm only works well in ideal datasets such as D1, and it is easy to be misled by the OOD actions in D2.
\our-CQL is the default setting of \our, which successfully learns a robust strategy with low exploitability in both D1 and D2 datasets.

\section{Supplement Results}
\label{app:more_results}

\subsection{Learning Curves of Nash Conv for Extensive-form Games}

\label{app:nc_curves}
\begin{figure} 

    \includegraphics[width=0.24\linewidth]{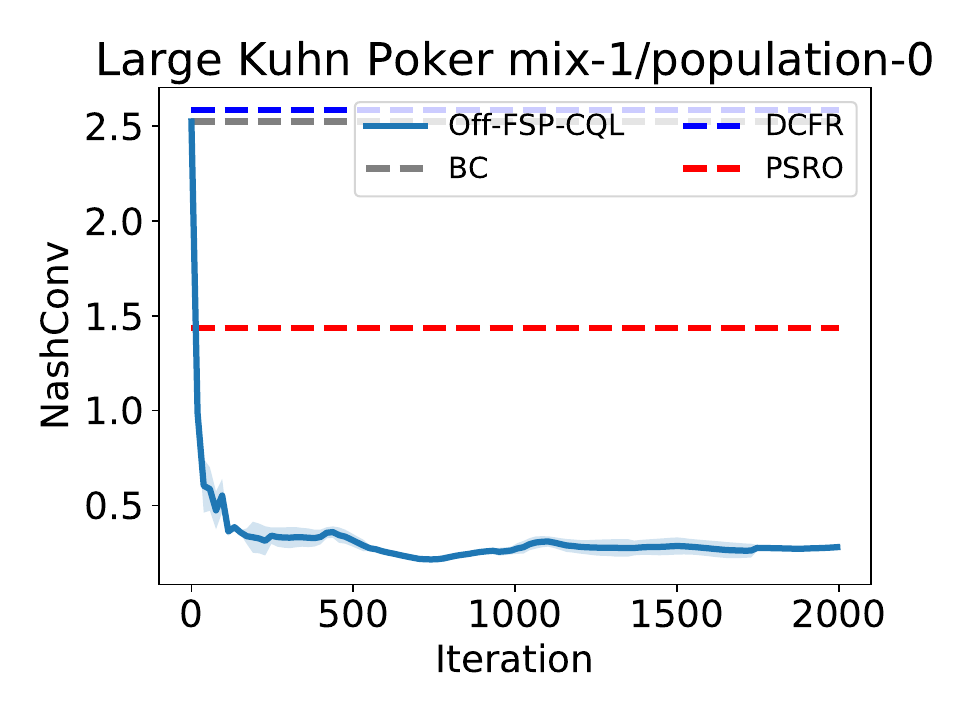}
    \includegraphics[width=0.24\linewidth]{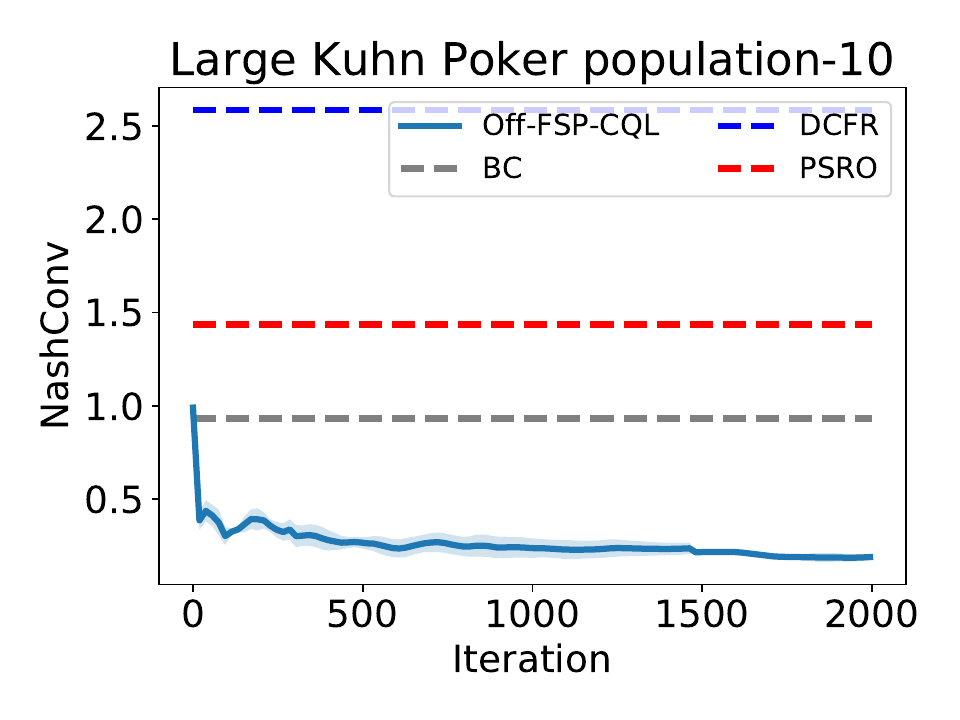}
    \includegraphics[width=0.24\linewidth]{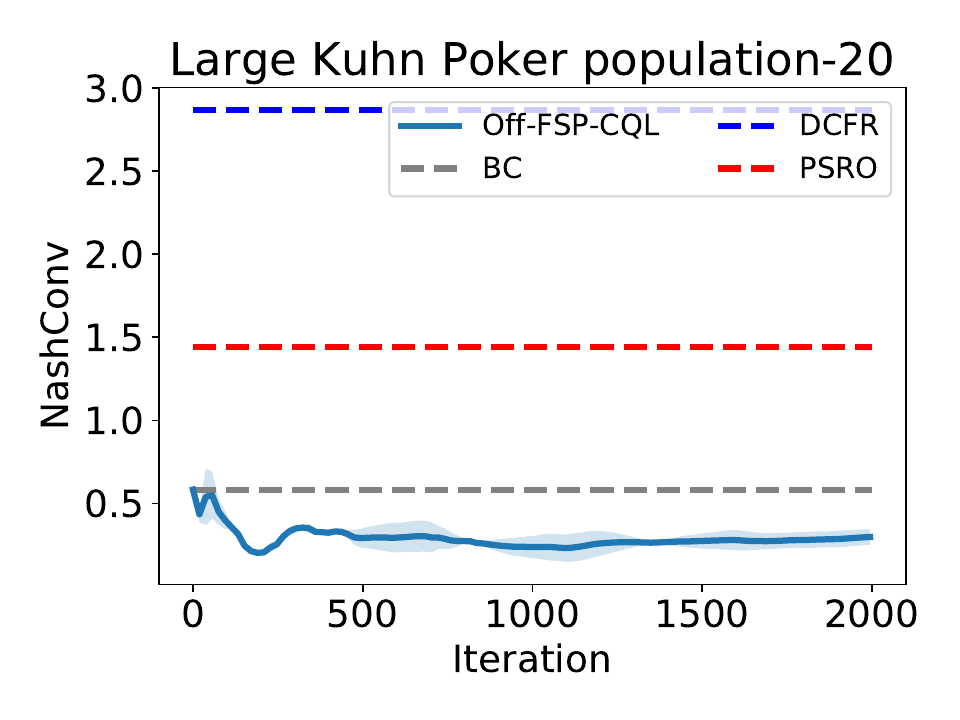}
    \includegraphics[width=0.24\linewidth]{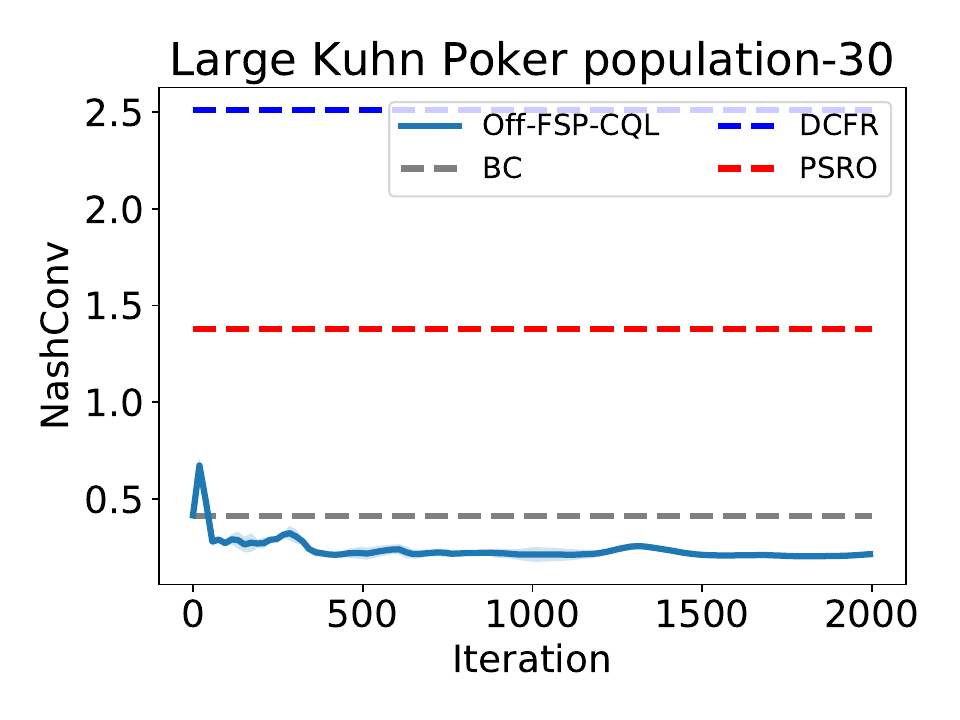}
    \includegraphics[width=0.24\linewidth]{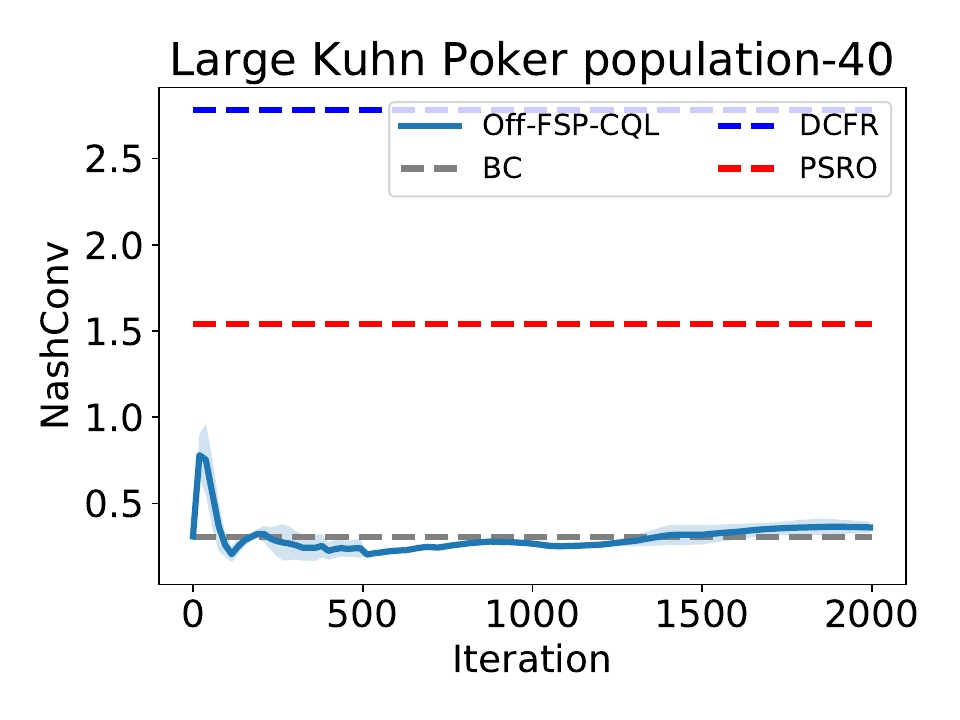}
    \includegraphics[width=0.24\linewidth]{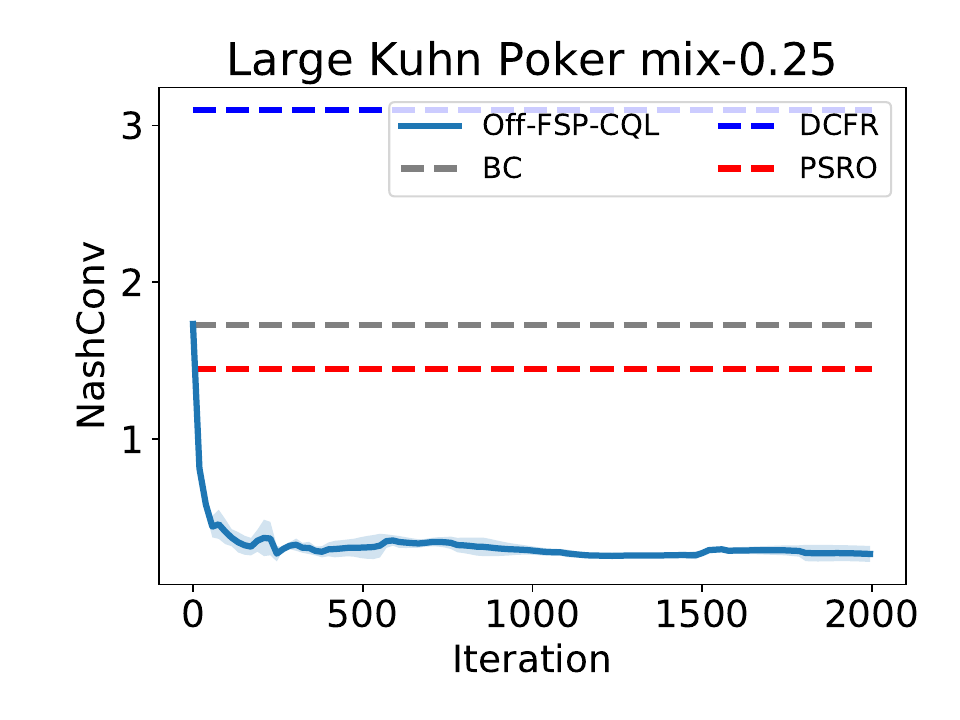}
    \includegraphics[width=0.24\linewidth]{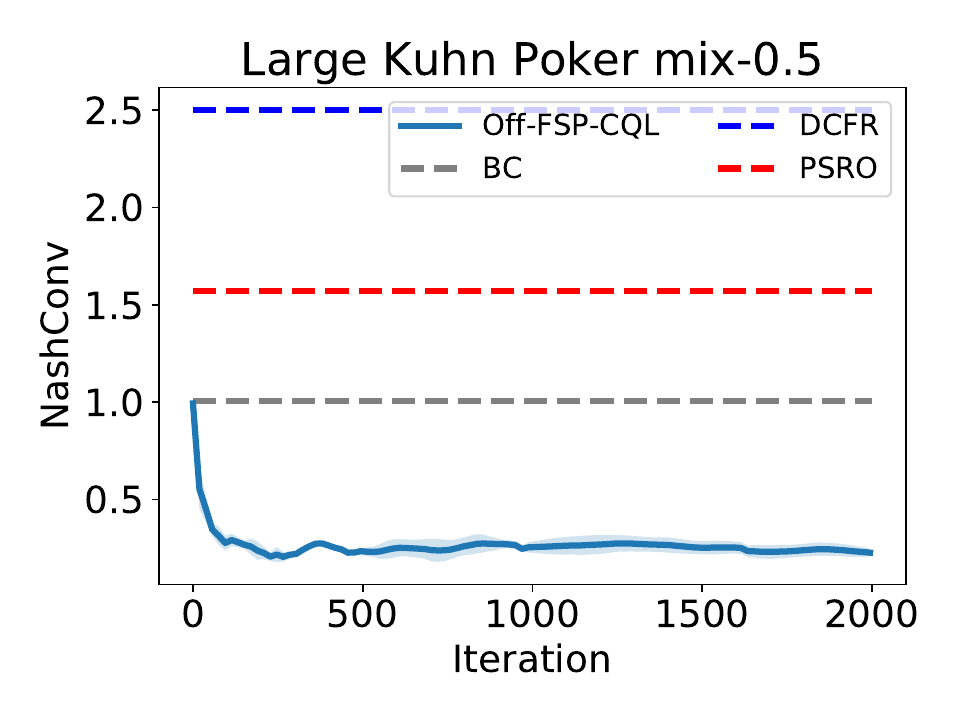}
    \includegraphics[width=0.24\linewidth]{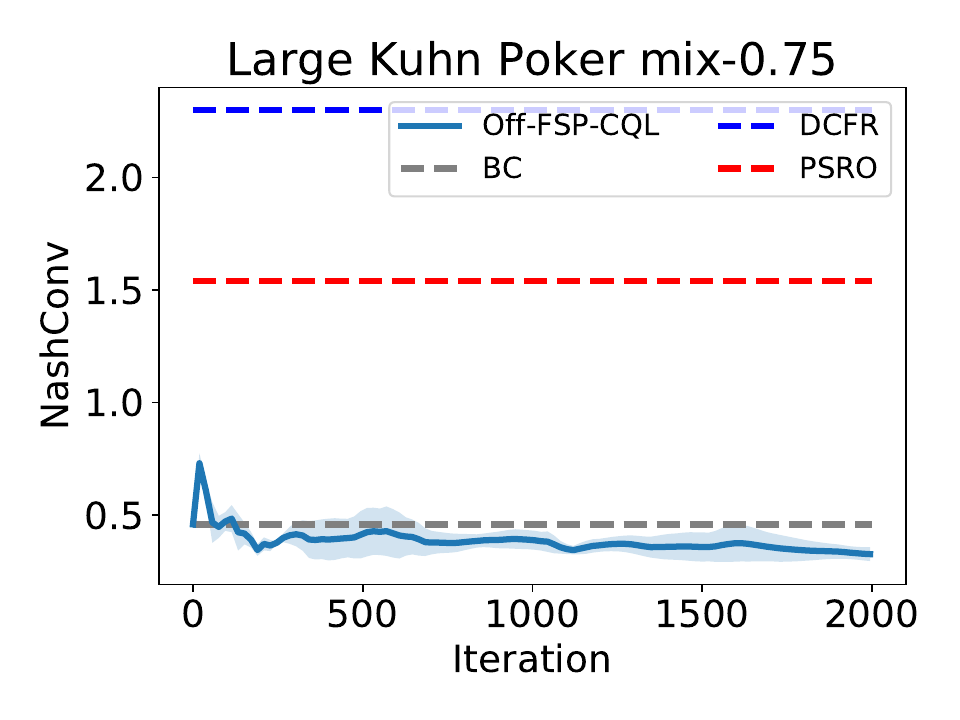}
    \includegraphics[width=0.24\linewidth]{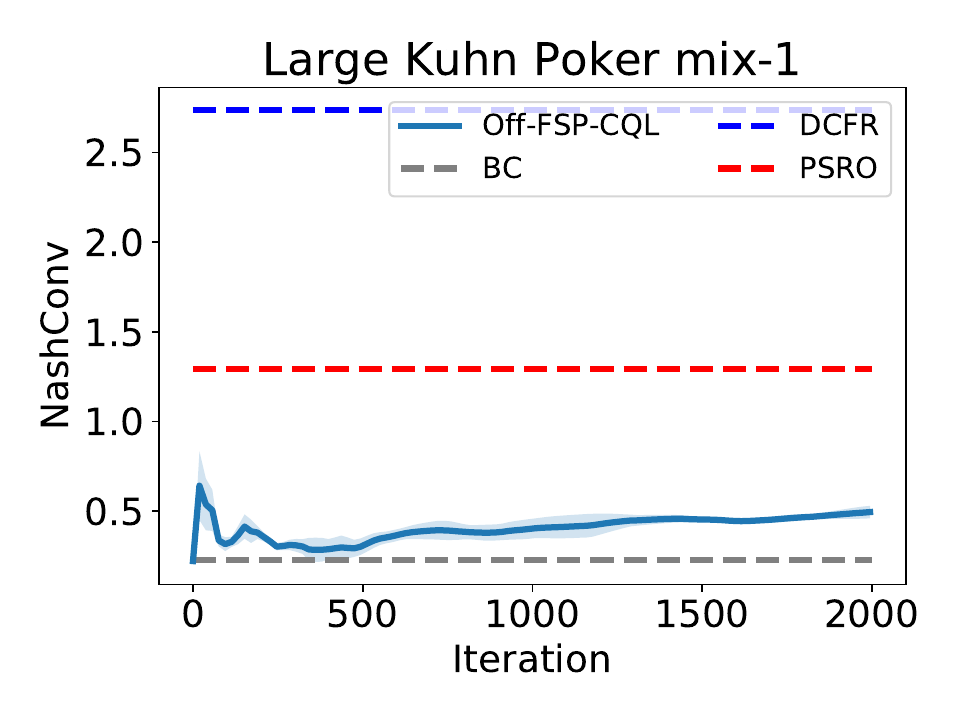}
    \includegraphics[width=0.24\linewidth]{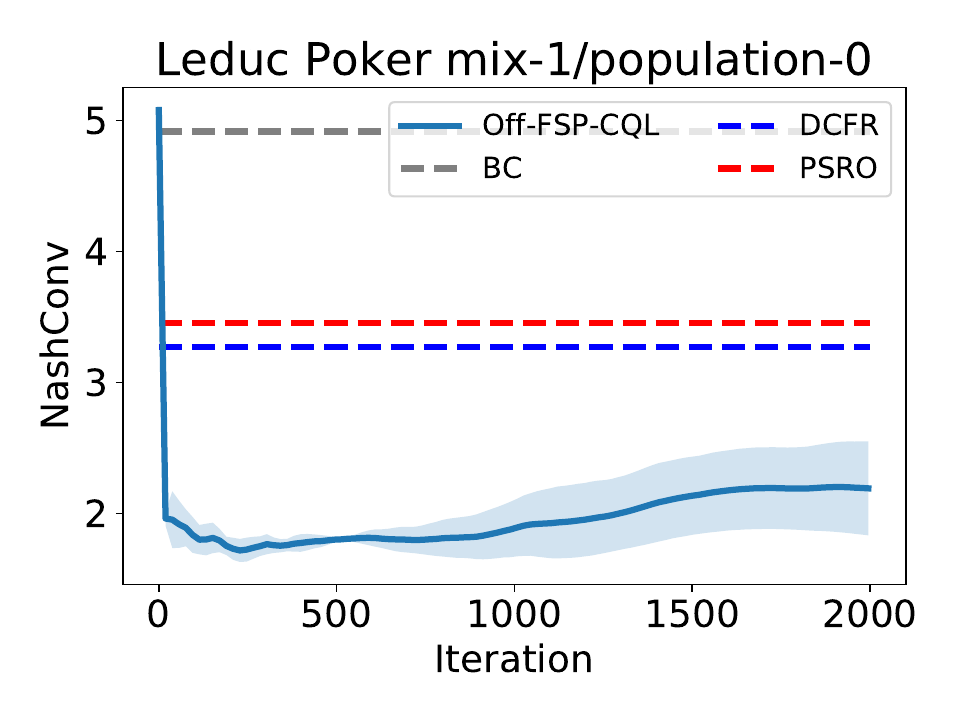}
    \includegraphics[width=0.24\linewidth]{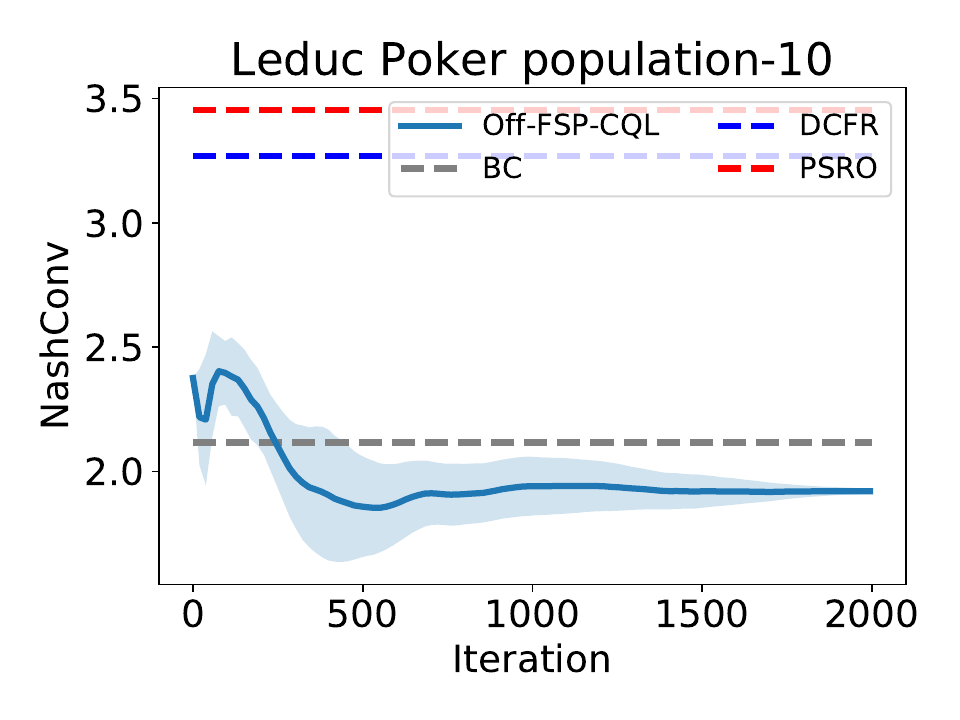}
    \includegraphics[width=0.24\linewidth]{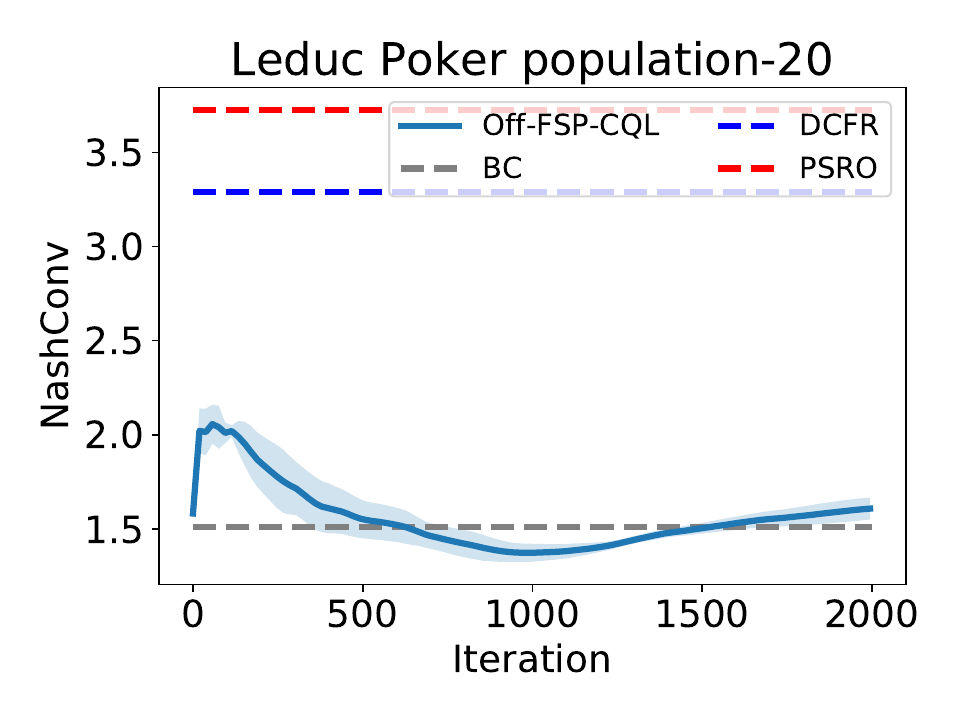}
    \includegraphics[width=0.24\linewidth]{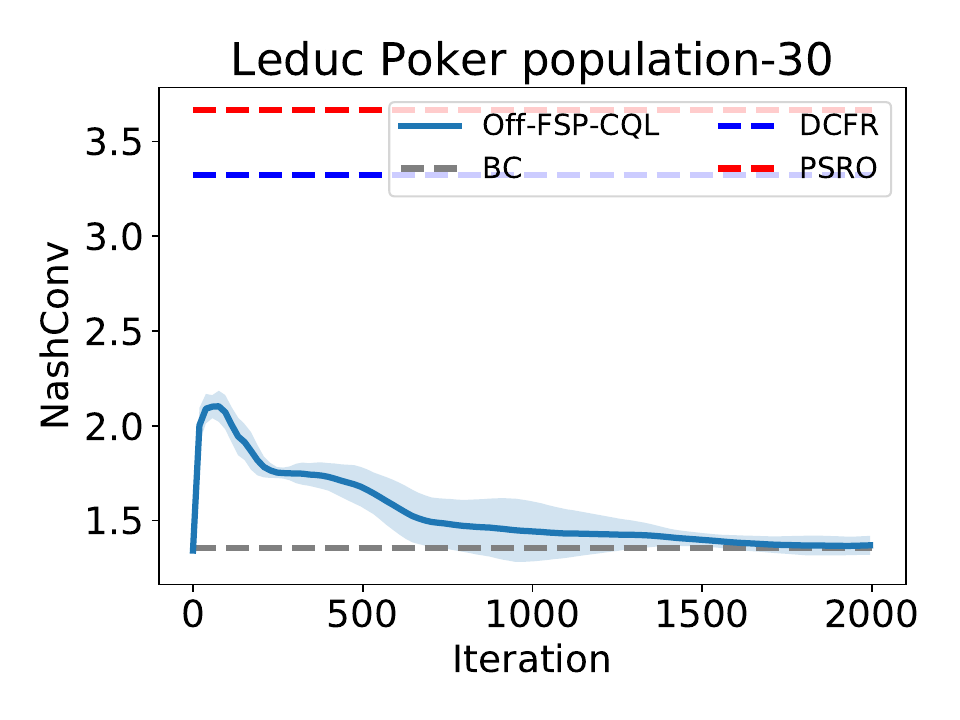}
    \includegraphics[width=0.24\linewidth]{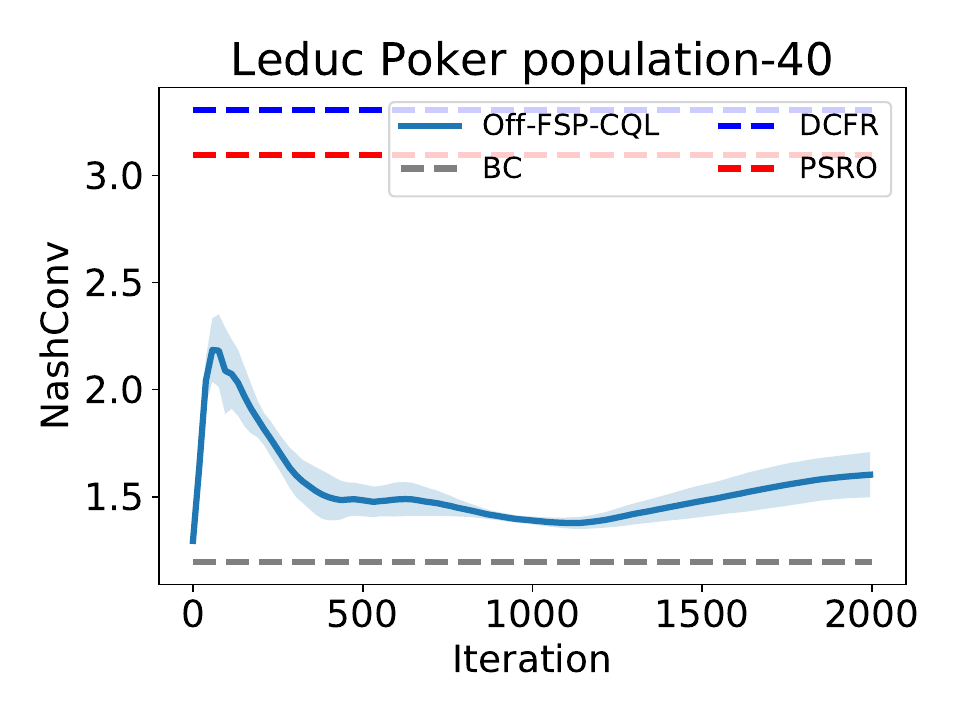}
    \includegraphics[width=0.24\linewidth]{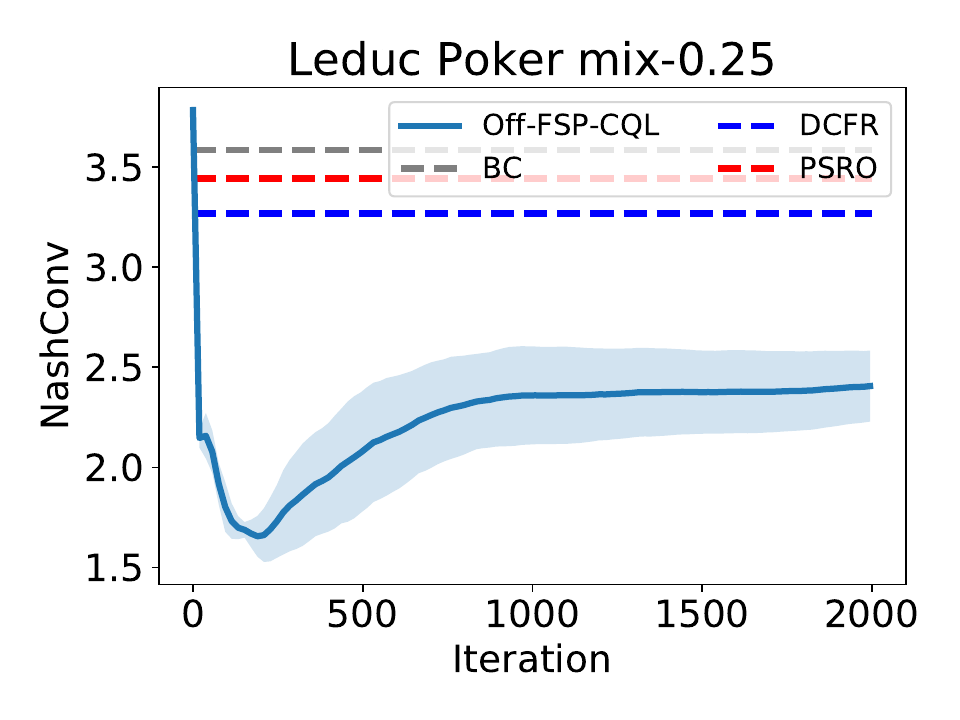}
    \includegraphics[width=0.24\linewidth]{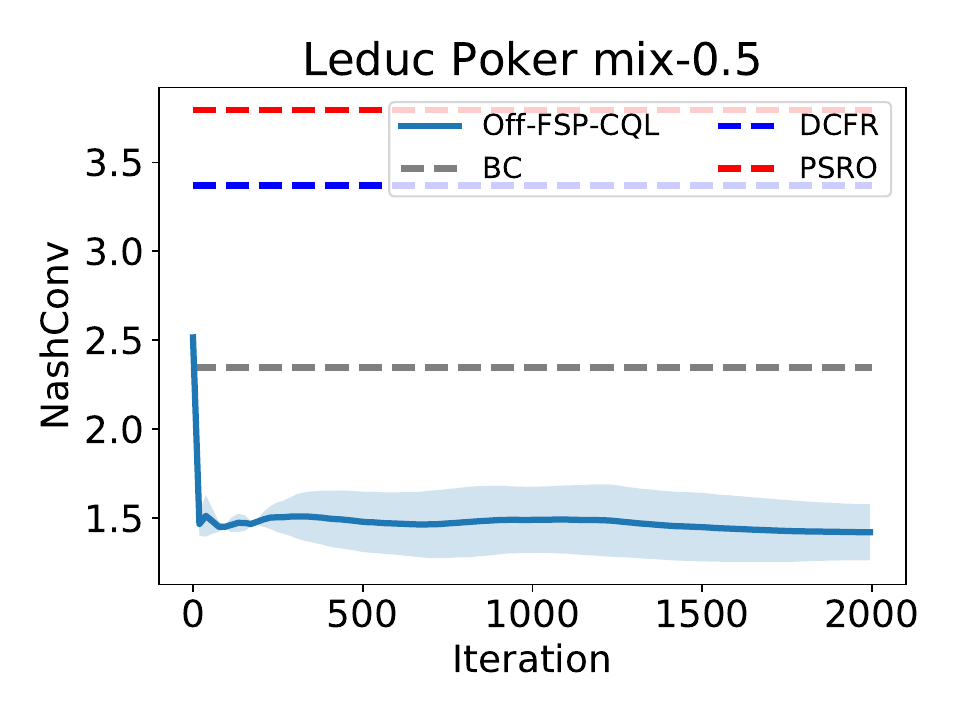}
    \includegraphics[width=0.24\linewidth]{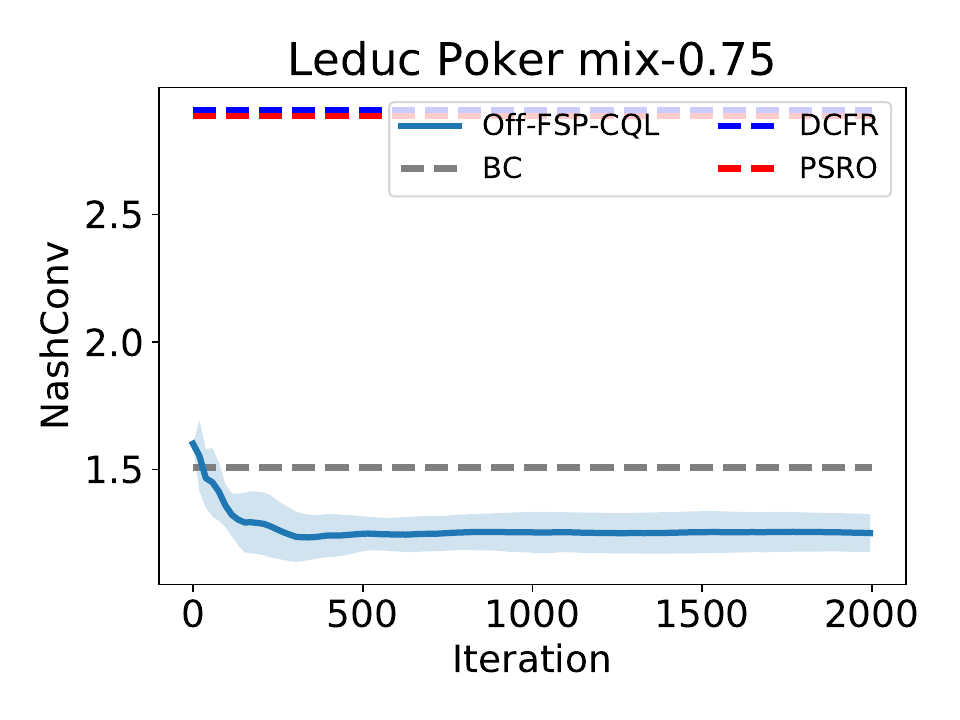}
    \includegraphics[width=0.24\linewidth]{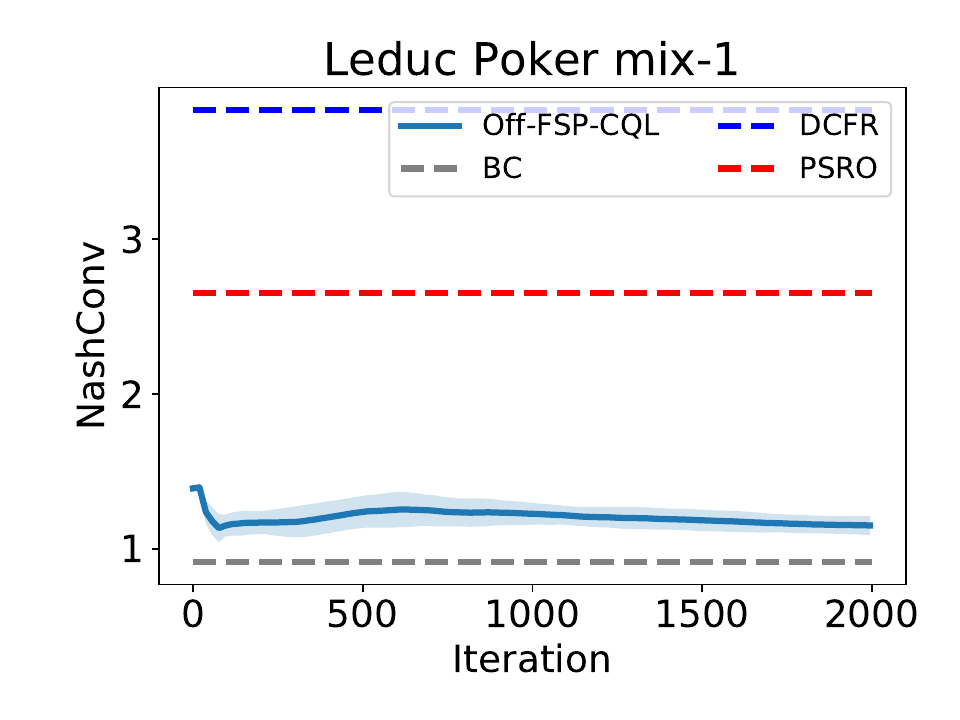}
    \includegraphics[width=0.24\linewidth]{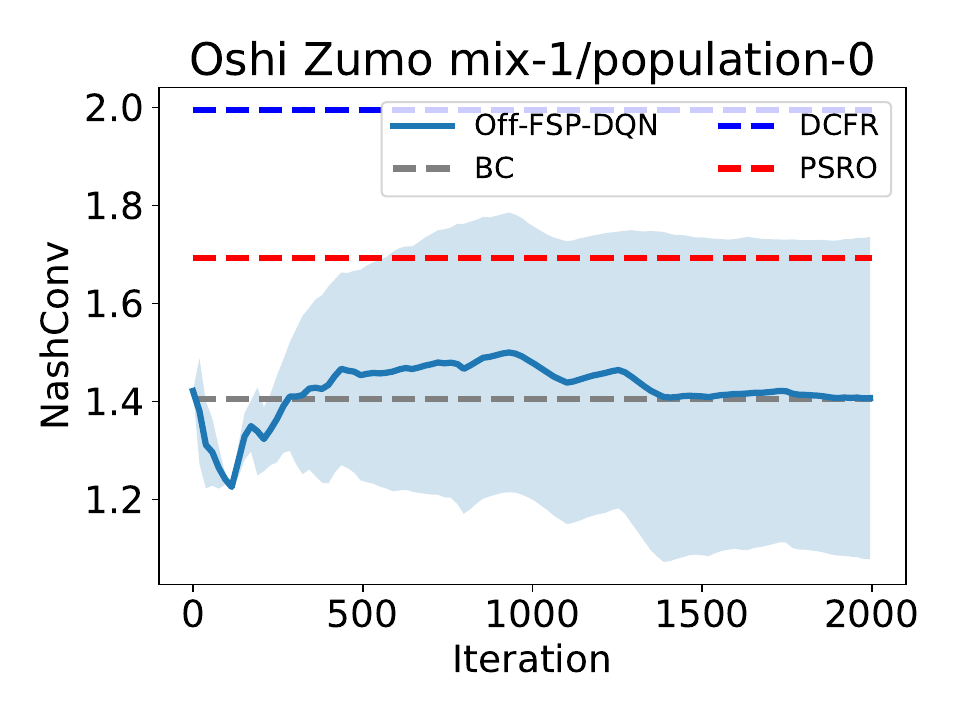}
    \includegraphics[width=0.24\linewidth]{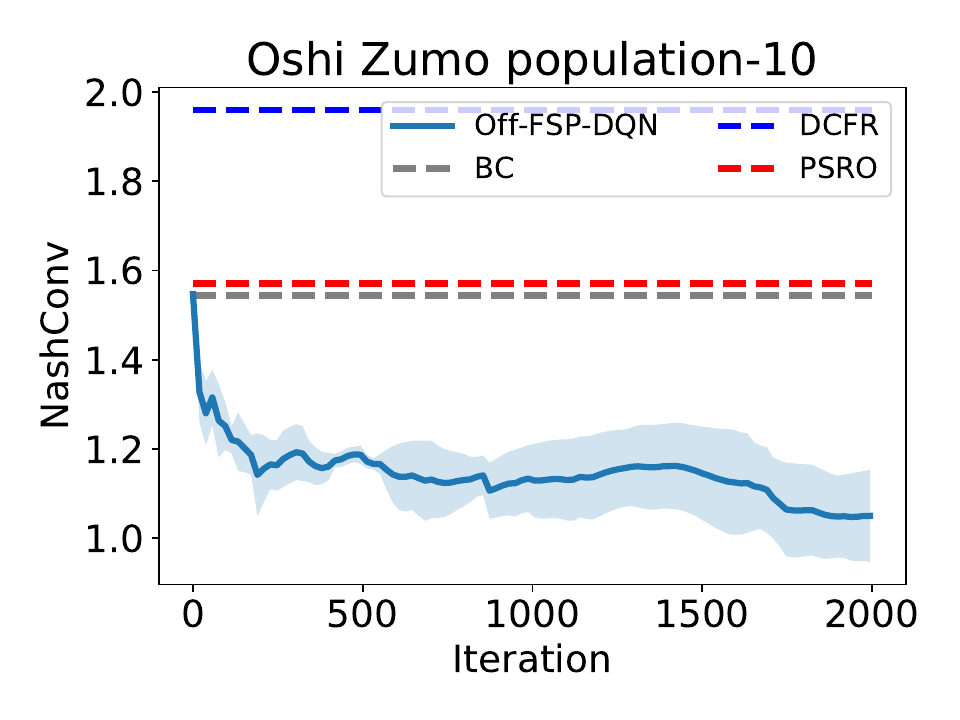}
    \includegraphics[width=0.24\linewidth]{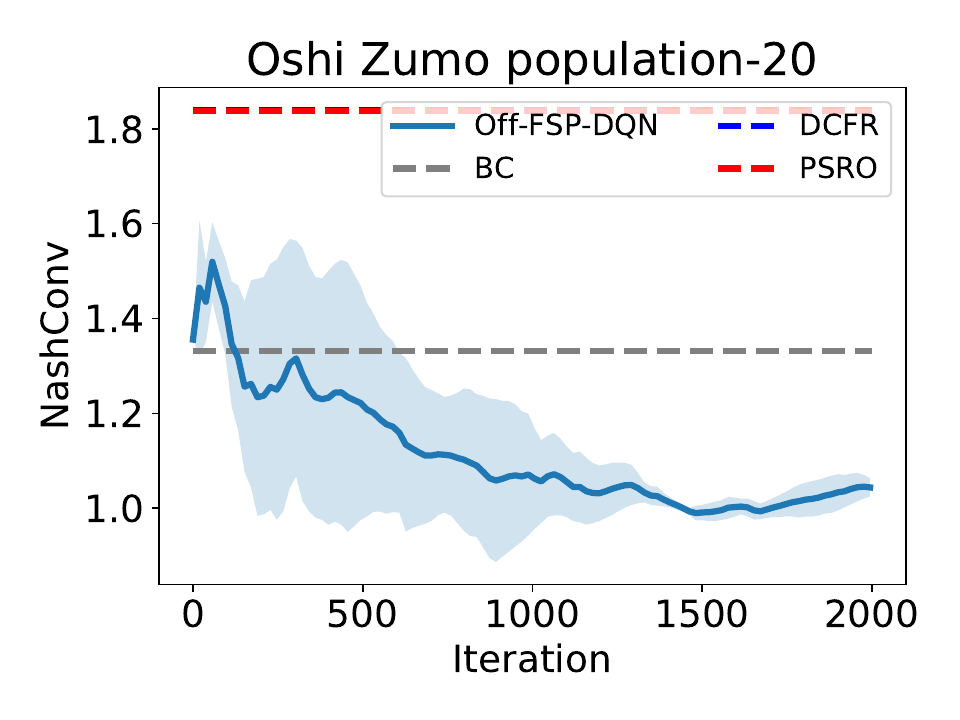}
    \includegraphics[width=0.24\linewidth]{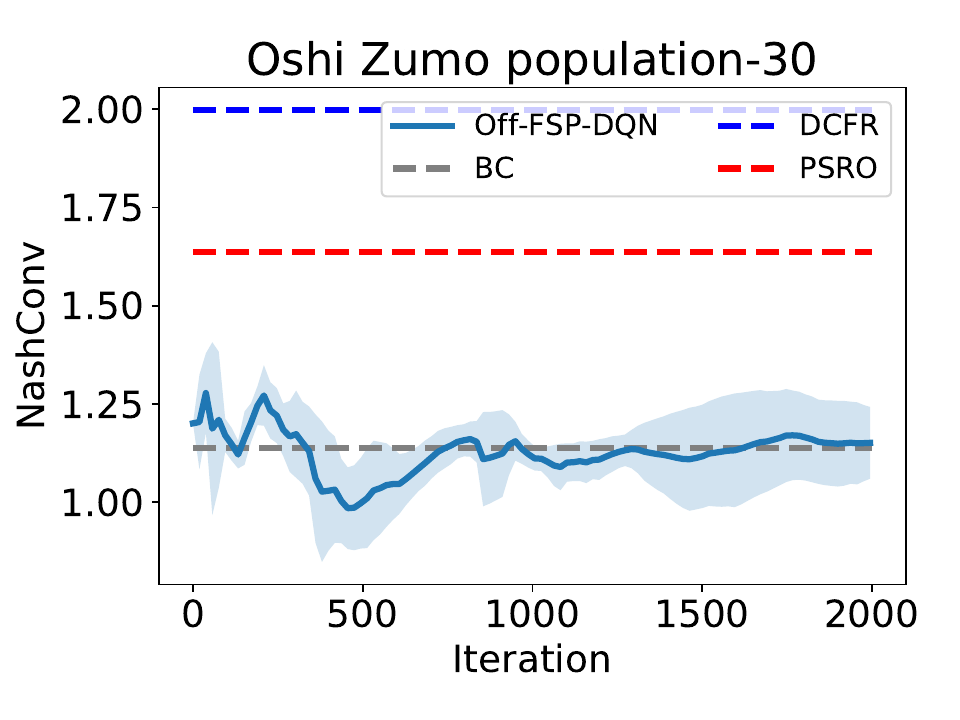}
    \includegraphics[width=0.24\linewidth]{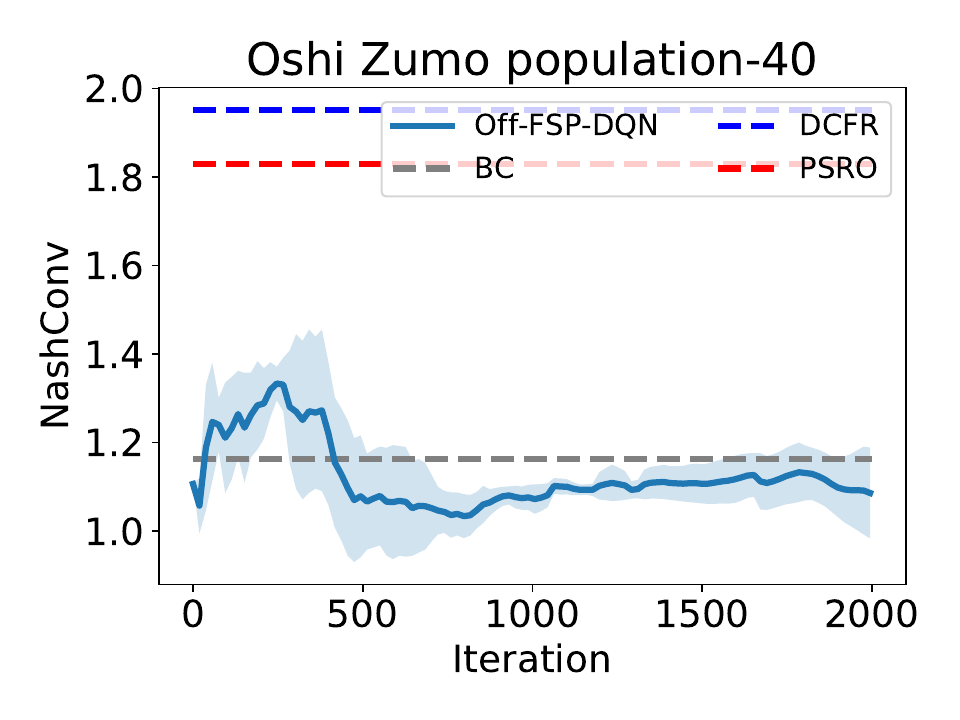}
    \includegraphics[width=0.24\linewidth]{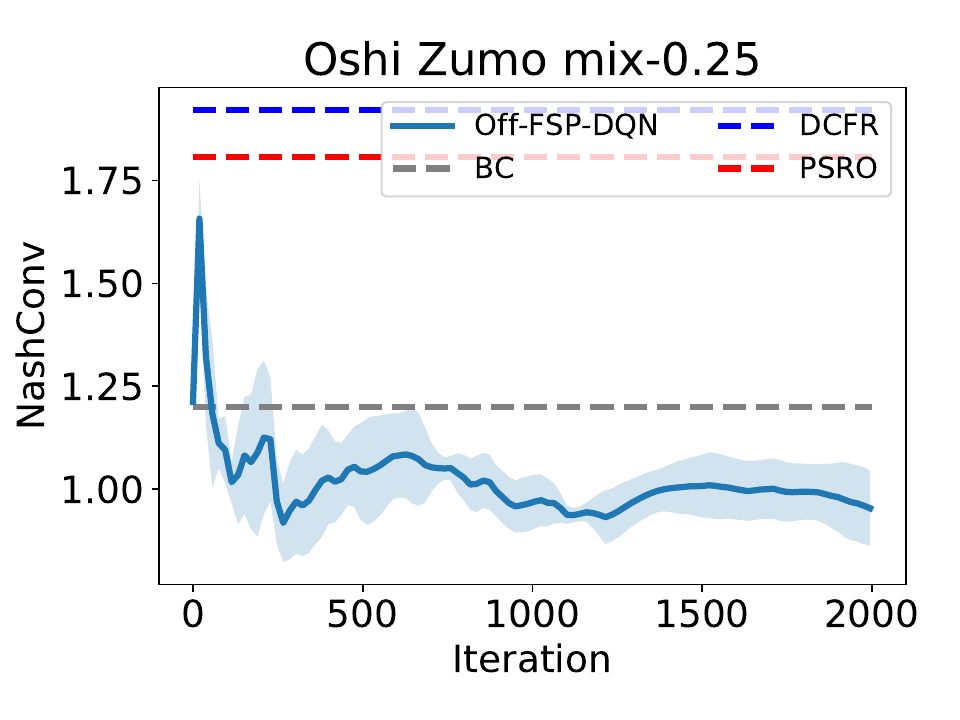}
    \includegraphics[width=0.24\linewidth]{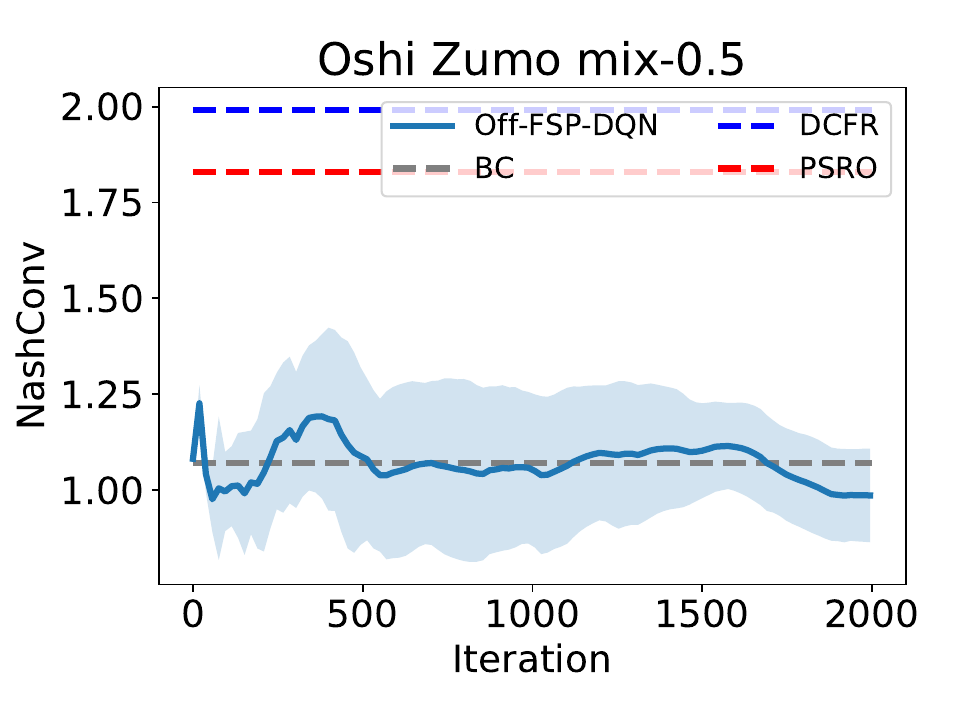}
    \includegraphics[width=0.24\linewidth]{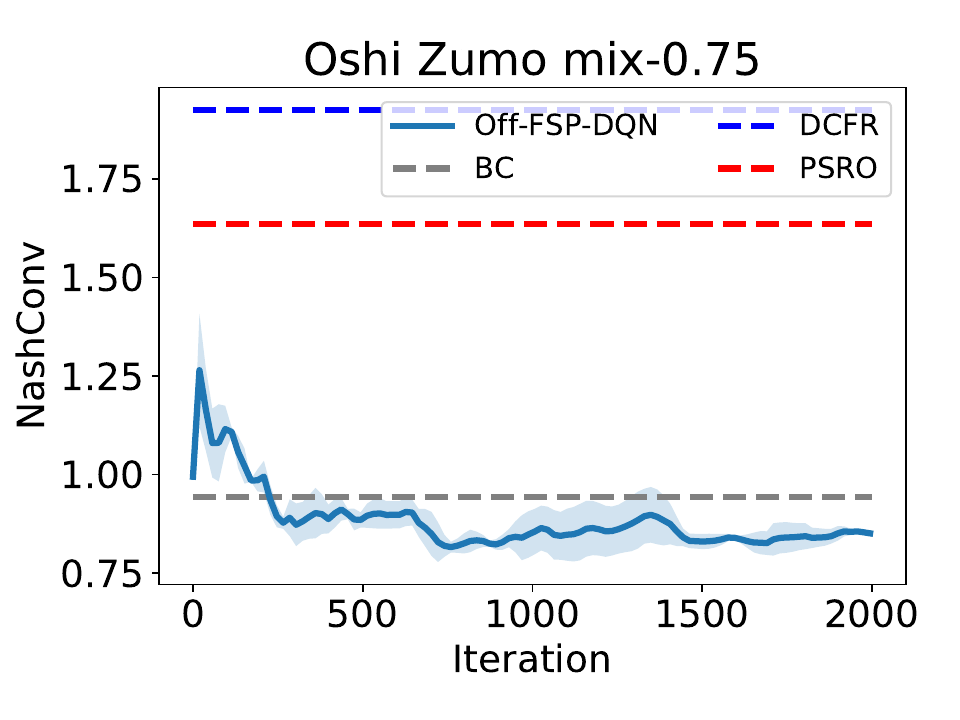}
    \includegraphics[width=0.24\linewidth]{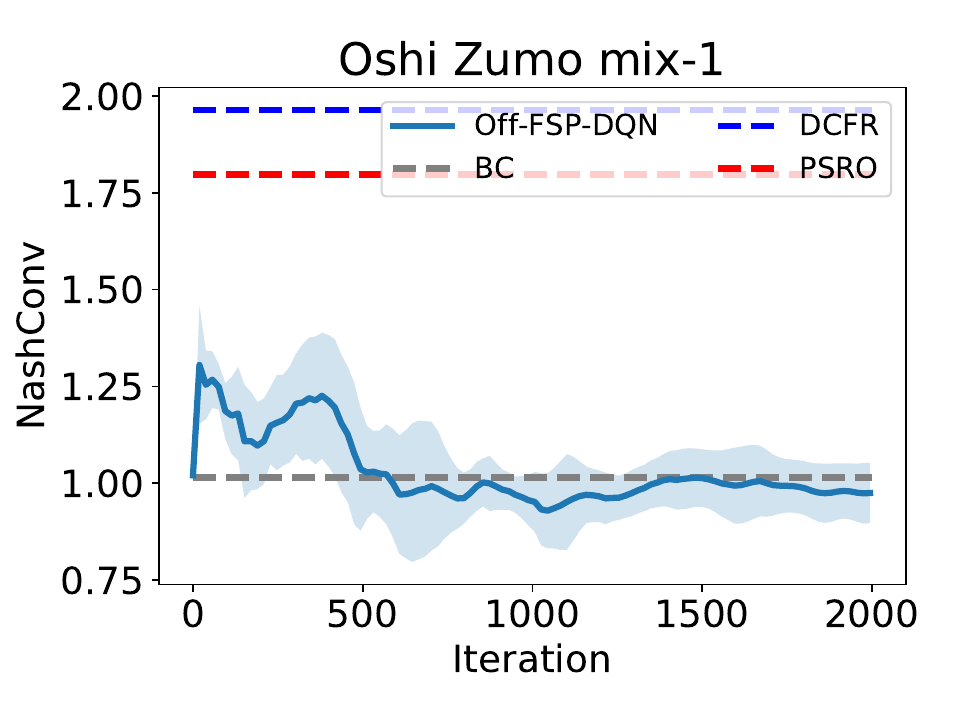}
    \caption{Learning Curves of our main experiments.}
    \label{fig:nc_curves_more}
\end{figure}
\Cref{fig:nc_curves_more} shows all the Nash Conv's learning curves of the main experiments in \Cref{sec:exp_leduc}.
 CQL is Q-learning algorithms with Q function only.

\begin{figure} \centering
    \begin{subfigure}{\linewidth}  
            \centering
        \includegraphics[width=0.32\linewidth]{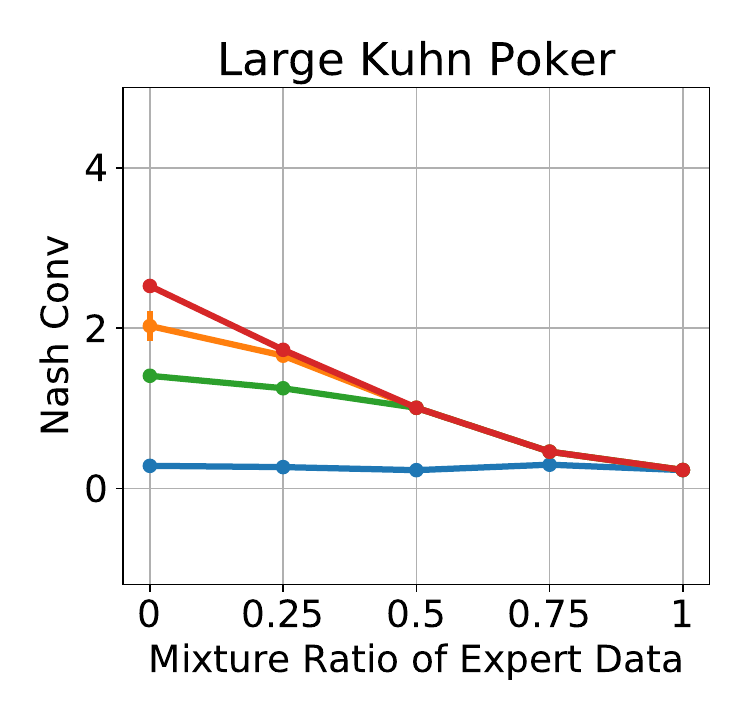}
        \includegraphics[width=0.32\linewidth]{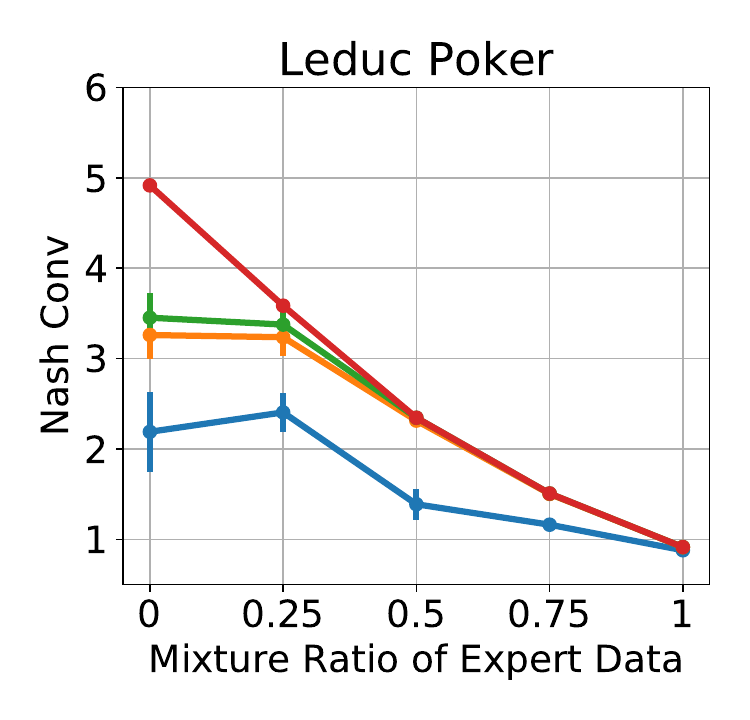}
        \includegraphics[width=0.32\linewidth]{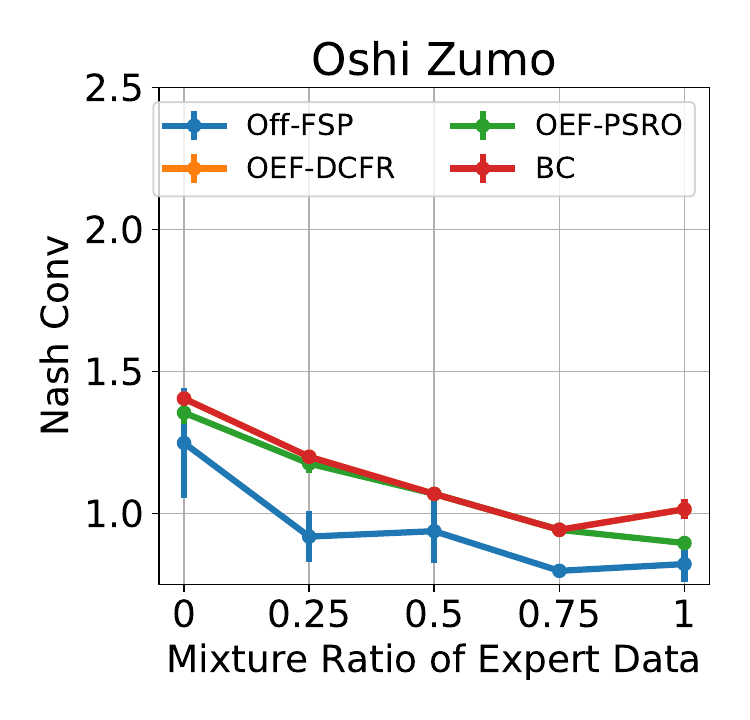}
        \\
        \includegraphics[width=0.32\linewidth]{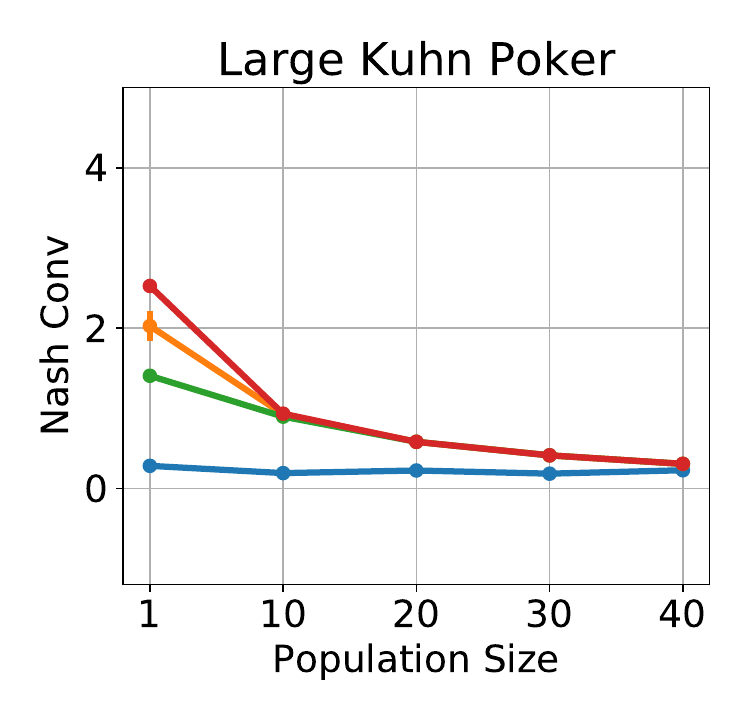}
        \includegraphics[width=0.32\linewidth]{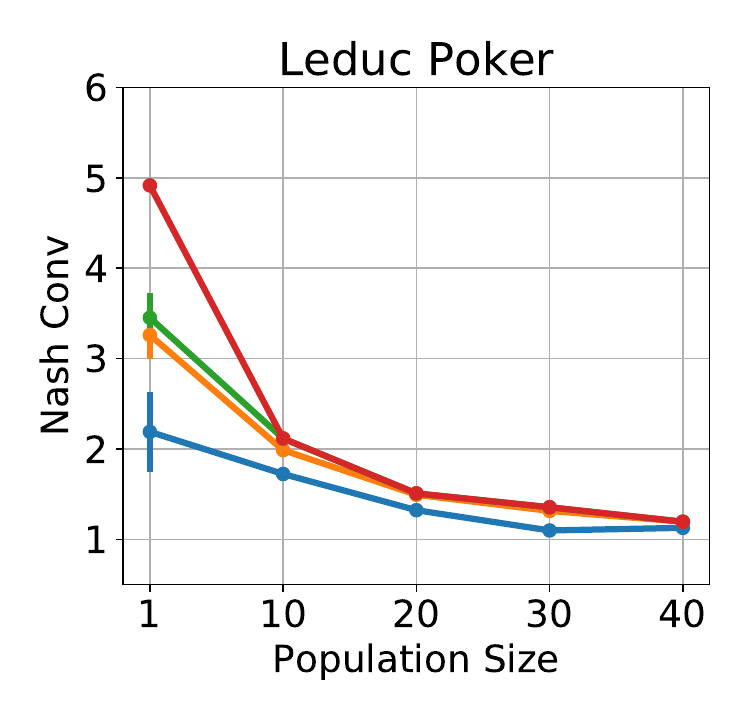}
        \includegraphics[width=0.32\linewidth]{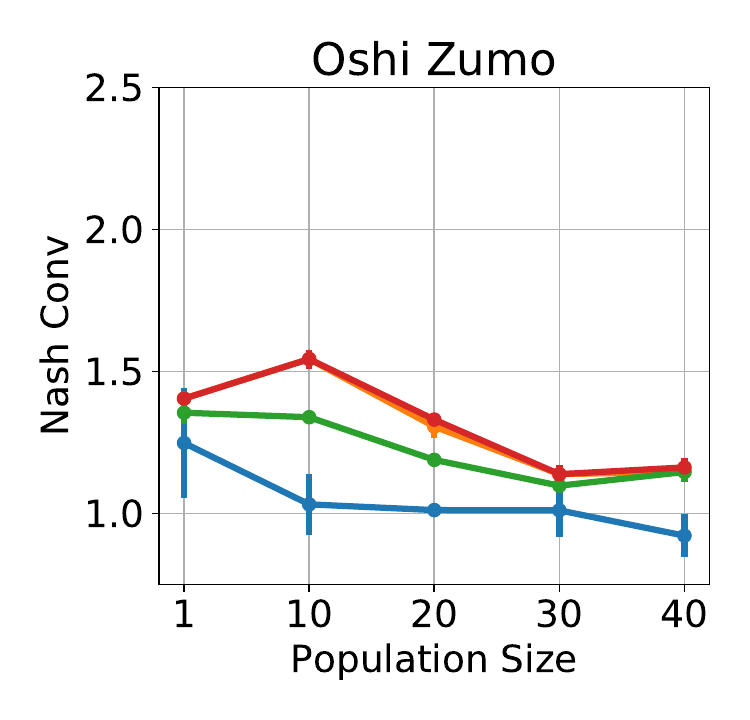}
        \label{fig:extgame_nc_oef}
        
    \end{subfigure} 
    \caption{Results on Extensive-Form Games with evaluation method in OEF. \textbf{(Top)} NashConv on Mix Datasets; \textbf{(Bottom)} NashConv on Population Datasets.}
    \label{fig:leduc_mix}
\end{figure}

\subsection{Experimental Results Under Rules of OEF}
OEF evaluates policies' performance by allowing policies to mix with BC by different weights. However, the results are decided by the best-mixed policies. 
This is equivalent to allowing policies to be interactively evaluated online, which is contrary to the setting of offline scenarios. It is difficult to figure out whether the original policy or BC causes this result.
We still provide the final results under this evaluation method in \Cref{fig:leduc_mix}, and our method still outperforms OEF.

\end{document}